\documentclass[aps,prd,preprint,groupedaddress]{revtex4}

\usepackage{epsfig}
\usepackage{graphics}

\begin{document}


\boldmath
\title{\vskip-3cm{\baselineskip14pt
\centerline{\normalsize\rm DESY 06-019\hfill ISSN 0418-9833}
\centerline{\normalsize\rm hep-ph/0602179\hfill}
\centerline{\normalsize\rm February 2006\hfill}}
\vskip1.5cm
Charmonium Production at High Energy in the $k_T$-Factorization
Approach}
\unboldmath

\author{\firstname{B.A.} \surname{Kniehl}}
\email{kniehl@desy.de}

\author{\firstname{D.V.} \surname{Vasin}}
\email{dmitriy.vasin@desy.de}
\thanks{on leave from Department of Physics, Samara State University,
Ac.~ Pavlov St.~1, 443011 Samara, Russia}

\affiliation{{II.} Institut f\" ur Theoretische Physik, Universit\" at Hamburg,
Luruper Chaussee 149, 22761 Hamburg, Germany}

\author{\firstname{V.A.} \surname{Saleev}}
\email{saleev@ssu.samara.ru}
\affiliation{Department of Physics, Samara State University,
Ac.~ Pavlov St.~1, 443011 Samara, Russia}

\begin{abstract}
We study charmonium production at high-energy colliders (Tevatron, HERA, and
LEP2) in the framework of the $k_T$-factorization approach and the
factorization formalism of non-relativistic quantum chromodynamics at leading
order in the strong-coupling constant $\alpha_s$ and the relative velocity $v$.
The transverse-momentum distributions of direct and prompt $J/\psi$-meson
production measured at the Fermilab Tevatron are fitted to obtain the
non-perturbative long-distance matrix elements for different choices of
unintegrated gluon distribution functions in the proton.
Using the matrix elements thus obtained, we predict charmonium production
rates in $\gamma\gamma$, $\gamma p$, and deep-inelastic $ep$ collisions
including the contributions from both direct and resolved photons.
The results are compared with the known ones obtained in the conventional
parton model and with recent experimental data from HERA and LEP2.
\end{abstract}


\maketitle

\section{Introduction}

Charmonium production at high energies has provided a useful laboratory for
testing the high-energy limit of quantum chromodynamics (QCD) as well as the
interplay of perturbative and non-perturbative phenomena in QCD.
The factorization formalism of non-relativistic QCD (NRQCD) \cite{NRQCD} is a
theoretical framework for the description of heavy-quarkonium production and
decay.
The factorization hypothesis of NRQCD assumes the separation of the effects of
long and short distances in heavy-quarkonium production.
NRQCD is organized as a perturbative expansion in two small parameters, the
strong-coupling constant $\alpha_s$ and the relative velocity $v$ of the heavy
quarks.

The phenomenology of strong interactions at high energies exhibits a dominant
role of gluon interactions in quarkonium production.
In the conventional parton model \cite{PartonModel}, the initial-state gluon
dynamics is controlled by the Dokshitzer-Gribov-Lipatov-Altarelli-Parisi
(DGLAP) evolution equation \cite{DGLAP}.
In this approach, it is assumed that $S > \mu^2 \gg \Lambda_{\rm QCD}^2$,
where $\sqrt{S}$ is the invariant collision energy, $\mu$ is the typical
energy scale of the hard interaction, and $\Lambda_{\rm QCD}$ is the
asymptotic scale parameter.
In this way, the DGLAP evolution equation takes into account only one big
logarithm, namely $\ln(\mu/\Lambda_{\rm QCD})$.
In fact, the collinear approximation is used, and the transverse momenta of
the incoming gluons are neglected.

In the high-energy limit, the contribution from the partonic
subprocesses involving $t$-channel gluon exchanges to the total
cross section can become dominant. The summation of the large
logarithms $\ln(\sqrt{S}/\mu)$ in the evolution equation can then
be more important than the one of the $\ln(\mu/\Lambda_{\rm
QCD})$ terms. In this case, the non-collinear gluon dynamics is
described by the Balitsky-Fadin-Kuraev-Lipatov (BFKL) evolution
equation \cite{BFKL}. In the region under consideration, the
transverse momenta ($k_T$) of the incoming gluons and their
off-shell properties can no longer be neglected, and we deal with
reggeized $t$-channel gluons. The theoretical framework for this
kind of high-energy phenomenology is the so-called
$k_T$-factorization approach \cite{KTGribov,KTCollins}, which can
be based on effective quantum field theory implemented with the
non-abelian gauge-invariant action, as was suggested a few years
ago~\cite{KTLipatov}.

This paper is organized as follows. In Sec.~\ref{sec:two}, the
$k_T$-factorization approach is briefly reviewed and compared
with the collinear parton model. The NRQCD formalism applied to
heavy-quarkonium production is briefly recapitulated in
Sec.~\ref{sec:three}. In Sec.~\ref{sec:four}, we present in
analytic form the squared amplitudes for $S$- and $P$-wave
quarkonium production via the fusion of reggeized gluons at
leading order (LO) in $\alpha_s$ and $v$. In Sec.~\ref{sec:five},
we perform fits to the transverse-momentum ($p_T$) distributions
of inclusive charmonium production measured at the Fermilab
Tevatron to obtain numerical values for the non-perturbative
matrix elements (NMEs) of the NRQCD factorization formalism. In
Secs.~\ref{sec:six} and \ref{sec:seven}, we compare our
theoretical predictions with recent experimental data of
charmonium production in $\gamma\gamma$, $\gamma p$, and
deep-inelastic $ep$ scattering at the DESY HERA and CERN LEP2
colliders. Section~\ref{sec:eight} contains our conclusions.

\boldmath
\section{\label{sec:two}The $k_T$-factorization approach}
\unboldmath

In the phenomenology of strong interactions at high energies, it
is necessary to describe the QCD evolution of the gluon
distribution functions of the colliding particles starting from
some scale $\mu_0$, which controls the non-perturbative regime,
to the typical scale $\mu$ of the hard-scattering processes,
which is typically of the order of the transverse mass
$M_T=\sqrt{M^2+|{\bf p}_T|^2}$ of the produced particle (or
hadron jet) with (invariant) mass $M$ and transverse two-momentum
${\bf p}_T$. In the region of very high energies, the typical
ratio $x=\mu/\sqrt{S}$ becomes very small, $x\ll1$. This leads to
large logarithmic contributions of the type $[\alpha_s
\ln(1/x)]^n$, which need to be resummed. This is conveniently
done by adopting the high-energy factorization scheme, which also
known as the $k_T$-factorization approach, in which the incoming
$t$-channel gluons have a finite transverse two-momentum ${\bf
k}_T$ and are off mass shell. This implies the notion of an
unintegrated gluon distribution function $\Phi(x,|{\bf
k}_T|^2,\mu^2)$. The resummation is then implemented by the BFKL
evolution equation \cite{BFKL}.

Effective Feynman rules for processes involving incoming off-shell gluons were
provided in Ref.~\cite{KTCollins}.
The special trick is to choose the polarization four-vector of the incoming
gluon as
\begin{equation}
\varepsilon^\mu(k_T)=\frac{k_T^\mu}{|{\bf
k}_T|},\label{eq:pol}
\end{equation}
where $k_T^\mu=(0,{\bf k}_T,0)$ is the transverse four-momentum of the gluon.
In the case of gluon-gluon fusion, the four-momenta of the incoming gluons can
be written as
\begin{eqnarray}
k_1^\mu=x_1P_1^\mu+k_{1T}^\mu,
\nonumber\\
k_2^\mu=x_2P_2^\mu+k_{2T}^\mu,
\end{eqnarray}
where $P_1^\mu=(\sqrt{S}/2)(1,0,0,1)$ and $P_2^\mu=(\sqrt{S}/2)(1,0,0,-1)$ are
the four-momenta of the colliding protons in the center-of-mass frame.
In the following, we shall also use the short-hand notation $p_T=|{\bf p}_T|$
etc.\ for the absolute of the transverse two-momentum.

In Ref.~\cite{KTLipatovFadin}, the incoming off-shell gluons are considered
as Reggeons (or reggeized gluons), which are interacting with quarks and
on-shell Yang-Mills gluons in a specific way.
Recently, in Ref.~\cite{KTAntonov}, the Feynman rules for the effective field
theory based on the non-abelian gauge-invariant action \cite{KTLipatov} were
derived for the vertices $RRg$, $Rgg$, $RRgg$, $Rggg$, and $RRggg$, where $R$
is an off-shell reggeized gluon and $g$ is an on-shell Yang-Mills gluon.
The interaction of a reggeized gluon with a quark is mediated via the
transition vertex $Rg$.
For the relevant LO amplitudes, which are calculated below, both approaches
\cite{KTCollins,KTLipatovFadin} give the same answers.
As was shown in Ref.~\cite{PRD2003}, the effective vertex $RRg$
\cite{KTLipatovFadin} can be obtained using the prescription
\cite{KTCollins} for the off-shell gluon polarization four-vector of
Eq.~(\ref{eq:pol}).

In the $k_T$-factorization approach, which is based on the high-energy limit
of QCD, the hadronic cross section of quarkonium (${\cal H}$) production
through the process
\begin{equation}
p + p \to {\cal H} + X \label{eq:ppHX}
\end{equation}
and the partonic cross section for the reggeized-gluon fusion subprocess
\begin{equation}
R + R \to {\cal H} + X\label{eq:RRHX}
\end{equation}
are related as
\begin{eqnarray}
\lefteqn{d\sigma^{\mathrm{KT}}(p + p \to {\cal H} + X, S)= \int{\frac{d
x_1}{x_1}} \int{d|{\bf k}_{1T}|^2}\int{\frac{d \varphi_1}{2
\pi}}\Phi(x_1,|{\bf k}_{1T}|^2,\mu^2)}
\nonumber\\
&\times&\int{\frac{d x_2}{x_2}} \int{d|{\bf
k}_{2T}|^2}\int{\frac{d \varphi_2}{2 \pi}}\Phi(x_2,|{\bf
k}_{2T}|^2,\mu^2) d\hat \sigma(R + R \to {\cal H} + X, {\bf
k}_{1T},{\bf k}_{2T}, \hat s), \label{eq:KT}
\end{eqnarray}
where $\hat s = x_1 x_2 S - ({\bf k}_{1T}+{\bf k}_{2T})^2$, $x_{1,2}$ are the
fractions of the proton momenta passed on to the reggeized gluons, and
$\varphi_{1,2}$ are the angles enclosed between ${\bf k}_{1,2T}$ and the
transverse momentum ${\bf p}_T$ of ${\cal H}$, which we take to point along
the $x$ axis.

In our numerical calculations, we use the unintegrated gluon distribution
functions by Bl\"umlein (JB) \cite{JB}, by Jung and Salam (JS) \cite{JS}, and
by Kimber, Martin, and Ryskin (KMR) \cite{KMR}.
A direct comparison between different unintegrated gluon distributions as
functions of $x$, $|{\bf k}_T|^2$, and $\mu^2$ may be found in
Ref.~\cite{PLB2002}.
Note, that the JB version is based on the BFKL evolution equation \cite{BFKL}.
On the contrary, the JS and KMR versions were obtained using the more
complicated Catani-Ciafaloni-Fiorani-Marchesini (CCFM) evolution equation
\cite{CCFM}, which takes into account both large logarithms of the types
$\ln(1/x)$ and $\ln(\mu/\Lambda_{\rm QCD})$.

For $\mu\gg \Lambda_{\rm QCD}$ and not too small $x=\mu/\sqrt{S}$, the
collinear approximation of the conventional parton model is recovered.
In the collinear parton model, the hadronic cross section
$d\sigma(p + p \to {\cal H} + X, S)$ and the relevant partonic cross section
$d\hat \sigma(g + g \to {\cal H} + X, \hat s)$ are related as
\begin{equation}
d\sigma^{\rm PM}(p + p \to {\cal H} + X, S)=\int dx_1\, G(x_1,\mu^2) \int
dx_2\, G(x_2,\mu^2)d \hat \sigma(g + g \to {\cal H} + X,\hat s),
\label{eq:PM}
\end{equation}
where $\hat s=x_1 x_2 S$ and $G(x,\mu^2)$ is the collinear gluon
distribution function of the proton, which satisfies the DGLAP \cite{DGLAP}
evolution equation.
The collinear and the unintegrated gluon distribution functions are formally
related as
\begin{equation}
x G(x,\mu^2)=\int_0^{\mu^2}{d|{\bf k}_{T}|^2}\,\Phi(x,|{\bf
k}_{T}|^2,\mu^2),
\end{equation}
so that the normalizations of Eqs.~(\ref{eq:KT}) and (\ref{eq:PM}) agree.

\section{\label{sec:three}NRQCD formalism}

In the framework of the NRQCD factorization approach \cite{NRQCD}, the
cross section of heavy-quarkonium production via a partonic subprocess
$a + b \to {\cal H} + X$ may be presented as a sum of terms in
which the effects of long and short distances are factorized as
\begin{equation}
d\hat \sigma (a + b \to {\cal H} + X)=\sum_n
d\hat \sigma (a + b \to Q\bar Q[n] + X)\langle{\cal O}^{\cal H}[n]\rangle,
\end{equation}
where $n$ denotes the set of color, spin, orbital and total angular momentum
quantum numbers of the $Q\bar Q$ pair and the four-momentum of the latter is
assumed to be equal to the one of the physical quarkonium state ${\cal H}$.
The cross section $d\hat \sigma (a + b \to Q\bar Q[n] + X)$ can be calculated
in perturbative QCD as an expansion in $\alpha_s$ using the non-relativistic
approximation for the relative motion of the heavy quarks in the $Q\bar Q$
pair.
The non-perturbative transition of the $Q\bar Q$ pair into ${\cal H}$ is
described by the NMEs $\langle {\cal O}^{\cal H}[n]\rangle$, which can be
extracted from experimental data.

To LO in $v$, we need to include the $c\bar c$ Fock states
$n = {^3S}_1^{(1)}, {^3S}_1^{(8)}, {^1S}_0^{(8)}, {^3P}_J^{(8)}$ if
${\cal H} = J/\psi, \psi^\prime$ and
$n = {^3P}_J^{(1)}, {^3S}_1^{(8)}$ if ${\cal H} = \chi_{cJ}$, where $J=0,1,2$.
Their NMEs satisfy the multiplicity relations
\begin{eqnarray}
\langle{\cal
O}^{\psi(nS)}[^3P_J^{(8)}]\rangle&=&(2J+1)\langle{\cal
O}^{\psi(nS)}[^3P_0^{(8)}]\rangle,\nonumber\\
\langle{\cal
O}^{\chi_{cJ}}[^3P_J^{(1)}]\rangle&=&(2J+1)\langle{\cal
O}^{\chi_{c0}}[^3P_0^{(1)}]\rangle,\nonumber\\
\langle{\cal
O}^{\chi_{cJ}}[^3S_1^{(8)}]\rangle&=&(2J+1)\langle{\cal
O}^{\chi_{c0}}[^3S_1^{(8)}]\rangle,
\end{eqnarray}
which follow to LO in $v$ from heavy-quark spin symmetry.
For example, in the case of $J/\psi$ production, the wave function of the
physical orthocharmonium state can be presented as a superposition of the Fock
states:
\begin{eqnarray}
|J/\psi \rangle&=&{\cal O}(v^0)|c\bar c [^3S_1^{(1)}]\rangle+{\cal
O}(v^1)|c\bar c [^3P_J^{(8)}]g\rangle +{\cal
O}(v^2)|c\bar c [^3S_1^{(1,8)}]gg\rangle
\nonumber\\
&&{}+{\cal O}(v^2)|c\bar c
[^1S_0^{(8)}]g\rangle+\cdots,\label{eq:JPsi}
\end{eqnarray}
where we use usual spectroscopic notation for the angular-momentum quantum
numbers of the $Q\bar Q$ pair and the index in parentheses $(1,8)$ denotes the
color state, either color singlet or color octet.
The color-singlet model (CSM) \cite{CSM} only takes into account the first
term in Eq.~(\ref{eq:JPsi}), which is of order $v^0$.
In this case, the NME $\langle{\cal O}^{J/\psi}[^3 S_1^{(1)}]\rangle$ is
directly related to the $J/\psi$ wave function at the origin $\Psi(0)$, which
can be calculated in the framework of the quark potential model \cite{QPM}, as
\begin{equation}
\langle {\cal O}^{J/\psi}[^3S_1^{(1)}]\rangle=2N_c(2J+1)|\Psi (0)|^2,
\end{equation}
where $N_c=3$ and $J=1$.
Similarly, the color-singlet $P$-wave NME reads
\begin{equation}
\langle{\cal O}^{\chi_{cJ}}[^3
P_J^{(1)}]\rangle=2N_c(2J+1)|\Psi^\prime(0)|^2,
\end{equation}
where $\Psi^\prime(0)$ is the derivative of the $\chi_{cJ}$ wave function at
the origin.

In the general case, the partonic cross section of quarkonium production
receives from the $Q\bar Q$ Fock state $n={}^{2S+1}L_J^{(1,8)}$ the
contribution \cite{NRQCD,Maltoni}
\begin{equation}
d\hat \sigma (a + b \to Q\bar Q[{}^{2S+1}L_J^{(1,8)}] \to {\cal
H})=d\hat \sigma (a + b \to Q\bar
Q[^{2S+1}L_J^{(1,8)}])\frac{\langle {\cal O}^{\cal
H}[^{2S+1}L_J^{(1,8)}]\rangle}{N_\mathrm{col}N_\mathrm{pol}},
\end{equation}
where $N_\mathrm{col}=2 N_c$ for the color-singlet state,
$N_\mathrm{col}=N_c^2-1$ for the color-octet state, and $N_\mathrm{pol}=2J+1$.
The partonic cross section of $Q\bar Q$ production is defined as
\begin{equation}
d\hat\sigma(a + b \to Q\bar
Q[^{2S+1}L_J^{(1,8)}])=\frac{1}{I}\overline{|{\cal A}(a + b
\to Q\bar Q[^{2S+1}L_J^{(1,8)}])|^2}d\Phi,
\end{equation}
where $I$ is the flux factor of the incoming particles, which is taken as in
the collinear parton model \cite{KTCollins} (for example,
$I= 2 x_1 x_2 S$ for process~(\ref{eq:RRHX})),
${\cal A}(a + b \to Q\bar Q[^{2S+1}L_J^{(1,8)}])$ is the production amplitude,
the bar indicates average (summation) over initial-state (final-state)
spins and colors, and $d\Phi$ is the phase space volume of the outgoing
particles.
This convention implies that the cross section in the $k_T$-factorization
approach is normalized approximately to the cross section for on-shell gluons
when ${\bf k}_{1T}={\bf k}_{2T}={\bf0}$.

The production amplitude ${\cal A}(a + b \to Q\bar Q[^{2S+1}L_J^{(1,8)}])$ can
be obtained from the one for an unspecified $Q\bar Q$ state,
${\cal A}(a + b \to Q\bar Q)$, by the application of appropriate projectors.
The projectors on the spin-zero and spin-one states read \cite{Guberina}:
\begin{eqnarray}
\Pi_0&=&\frac{1}{\sqrt{8m^3}}\left(\frac{\hat p}{2}-\hat
q-m\right)\gamma_5\left(\frac{\hat p}{2}+\hat q+m\right),
\nonumber\\
\Pi_1^\alpha&=&\frac{1}{\sqrt{8m^3}}\left(\frac{\hat p}{2}-\hat
q-m\right)\gamma^\alpha\left(\frac{\hat p}{2}+\hat
q+m\right),
\end{eqnarray}
respectively, where $\hat p=\gamma^\mu p_\mu$, $p^\mu$ is the four-momentum of
the $Q\bar Q$ pair, $q^\mu$ is the four-momentum of the relative motion,
$m=M/2$ is the mass of the quark $Q$, and $M$ is the mass of the quarkonium
state ${\cal H}$.
In our numerical calculations, we use $m_c=1.55$ GeV.
The projection operators for the color-singlet and color-octet states read:
\begin{eqnarray}
C_1=\frac{\delta_{ij}}{\sqrt{N_c}},\nonumber\\
C_8=\sqrt{2}T^a_{ij},
\end{eqnarray}
respectively, where $T^a$ with $a=1,\ldots,N_c^2-1$ are the generators of the
color gauge group SU(3).
To obtain the projection on a state with orbital-angular-momentum quantum
number $L$, we need to take $L$ times the derivative with respect to $q$ and
then put $q=0$.
For the processes discussed here, we have
\begin{eqnarray}
{\cal A}(a + b \to Q\bar Q[^1S_0^{(1,8)}])&=&\mbox{Tr}\left[
C_{1,8}\Pi_0{\cal A}(a + b \to Q\bar Q)\right]_{q=0},
\nonumber\\
{\cal A}(a + b \to Q\bar Q [^3S_1^{(1,8)}])&=&\mbox{Tr}\left[
C_{1,8}\Pi_1^\alpha{\cal
A}(a + b \to Q\bar Q)\varepsilon_\alpha(p)\right]_{q=0},
\nonumber\\
{\cal A}(a + b \to Q\bar Q
[^3P_J^{(1,8)}])&=&\frac{d}{dq_\beta}\mbox{Tr}\left[
C_{1,8}\Pi_1^\alpha{\cal A}(a + b \to Q\bar Q
)\varepsilon_{\alpha\beta}(p)\right]_{q=0},
\end{eqnarray}
where $\varepsilon_\alpha(p)$ is the polarization four-vector of a spin-one
particle with four-momentum $p^\mu$ and mass $M=p^2$ and
$\varepsilon_{\alpha\beta}(p)$ is its counterpart for a spin-two particle.
For the ${}^3 S_1$ state, the polarization sum reads
\begin{equation}
\sum_{J_z}\varepsilon_\alpha(p)\varepsilon^*_{\alpha^\prime}(p)
={\cal P}_{\alpha\alpha^\prime}(p)
=-g_{\alpha\alpha^\prime}+\frac{p_\alpha p_{\alpha^\prime}}{M^2}.
\end{equation}
For the ${}^3 P_J$ states with $J=0,1,2$, we have
\begin{eqnarray}
\varepsilon^{(0)}_{\alpha\beta}(p)
\varepsilon^{(0)*}_{\alpha^\prime\beta^\prime}(p)&=&
\frac{1}{3}{\cal P}_{\alpha\beta}(p)
{\cal P}_{\alpha^\prime\beta^\prime}(p),
\nonumber\\
\sum_{J_z}\varepsilon^{(1)}_{\alpha\beta}(p)
\varepsilon^{(1)*}_{\alpha^\prime\beta^\prime}(p)&=&
\frac{1}{2}\left[{\cal P}_{\alpha\alpha^\prime}(p){\cal
P}_{\beta\beta^\prime}(p)-{\cal P}_{\alpha\beta^\prime}(p){\cal
P}_{\alpha^\prime\beta}(p)\right],
\nonumber\\
\sum_{J_z}\varepsilon^{(2)}_{\alpha\beta}(p)
\varepsilon^{(2)*}_{\alpha^\prime\beta^\prime}(p)&=&
\frac{1}{2}\left[{\cal P}_{\alpha\alpha^\prime}(p){\cal
P}_{\beta\beta^\prime}(p)+{\cal P}_{\alpha\beta^\prime}(p){\cal
P}_{\alpha^\prime\beta}(p)\right]-\frac{1}{3}{\cal P}_{\alpha\beta}{\cal
P}_{\alpha^\prime\beta^\prime}(p).
\end{eqnarray}

The subprocesses relevant for our analysis read:
$R + R\to Q\bar Q$, $R + R \to Q\overline{Q} + g$,
$R + \gamma \to Q\overline{Q}$, $R+ \gamma \to Q\overline{Q} + g$,
$R + e \to e + Q\overline{Q}$, and $R + e \to e + Q\overline{Q} + g$.

\section{\label{sec:four}Charmonium production by reggeized gluons}

In this section, we obtain the squared amplitudes for inclusive
charmonium production via the fusion of two reggeized gluons or a
reggeized gluon and a real or virtual photon in the framework of
NRQCD. We work at LO in $\alpha_s$ and $v$ and consider the
following partonic subprocesses:
\begin{eqnarray}
R + R &\to& {\cal H} [{^3P}_J^{(1)},{^3S}_1^{(8)},{^1S}_0^{(8)},{^3P}_J^{(8)}],
\label{eq:RRtoH}\\
R + R &\to& {\cal H} [{^3S}_1^{(1)}] + g,
\label{eq:RRtoHG}\\
R+\gamma &\to&  {\cal H} [{^3S}_1^{(8)},{^1S}_0^{(8)},{^3P}_J^{(8)}],
\label{eq:FRtoH}\\
R+\gamma &\to& {\cal H} [{^3S}_1^{(1)}] + g,
\label{eq:FRtoHG}\\
R + e &\to&  e + {\cal H} [{^3S}_1^{(8)},{^1S}_0^{(8)},{^3P}_J^{(8)}],
\label{eq:eRtoH}\\
R + e &\to& e + {\cal H} [{^3S}_1^{(1)}] + g.
\label{eq:eRtoHG}
\end{eqnarray}
Notice that, in the collinear parton model, subprocesses (\ref{eq:RRtoH}),
(\ref{eq:FRtoH}), and (\ref{eq:eRtoH}) only contribute for $p_T\approx 0$.
Therefore, to LO in the collinear parton model, we need to take into account
the corresponding subprocesses with an additional hard gluon in the final
state, for example $g + g \to {\cal H}[^3S_1^{(8)}] + g$.
The amplitudes of these color-octet subprocesses, after replacing $g\to R$ in
the initial state, are of next-to-leading order (NLO) in the
$k_T$-factorization approach and suffer from infrared divergences, in contrast
to the subprocesses (\ref{eq:RRtoHG}) and (\ref{eq:FRtoHG}) in the
color-singlet channel.
The analysis of NLO contributions to inclusive charmonium production by
reggeized gluon-gluon fusion in the $k_T$-factorization approach is beyond the
scope of this paper and needs a separate investigation.

The phenomenological procedure, adopted in Ref.~\cite{CDFBaranov}, to
regularize infrared divergences due to propagators getting on-shell with the
help of some cut parameter, which is unknown a priori, is likely to be
problematic.
The analysis of NLO corrections in the $k_T$-factorization approach is
currently an open issue, which has been consistently solved only in part,
{\it e.g.}\ in Ref. \cite{Ostrovsky}, where NLO corrections to the subprocess
$R + R \to g$ were studied.

According to the prescription of Ref.~\cite{KTCollins}, the amplitude of
$R + R \to c + \bar c (+ g)$ is related to the one of
$g + g \to c + \bar c (+ g)$ by
\begin{equation}
{\cal A}(R + R \to c + \bar c (+ g))=\varepsilon^\mu(k_1)
\varepsilon^\nu(k_2){\cal A}_{\mu\nu}(g + g \to c + \bar c (+g)),
\label{eq:RRcc}
\end{equation}
where $\varepsilon^\mu(k_1)$ and $\varepsilon^\mu(k_2)$ are defined according
to Eq.~(\ref{eq:pol}).
Analogous relations hold for $R + \gamma \to c + \bar c (+ g)$ and
$R + e\to e + c + \bar c (+ g)$.
The amplitudes of the relevant QCD subprocesses $g + g \to c + \bar c (+ g)$,
$g + \gamma \to c + \bar c (+ g)$, and $g + e \to e+ c + \bar c (+ g)$ are
evaluated using the conventional Feynman rules of QCD.

We now present and discuss our results for the squared amplitudes of
subprocesses~(\ref{eq:RRtoH}) and (\ref{eq:RRtoHG}), contributing to
hadroproduction.
In the case of the $2\to1$ subprocesses~(\ref{eq:RRtoH}), we obtain
\begin{eqnarray}
\overline{|{\cal A}(R + R \to {\cal H}[^3P_0^{(1)}]|^2}
&=&\frac{8}{3}\pi^2 \alpha_s^2\frac{\langle{\cal
O}^{\cal H}[^3P_0^{(1)}]\rangle}{M^5}
F^{[^3P_0]}(t_1,t_2,\varphi),
\nonumber\\
\overline{|{\cal A}(R + R \to {\cal H}[^3P_1^{(1)}]|^2}
&=&\frac{16}{3}\pi^2 \alpha_s^2\frac{\langle{\cal
O}^{\cal H}[^3P_1^{(1)}]\rangle}{M^5}
F^{[^3P_1]}(t_1,t_2,\varphi),
\nonumber\\
\overline{|{\cal A}(R + R \to {\cal H}[^3P_2^{(1)}]|^2}
&=&\frac{32}{45}\pi^2
\alpha_s^2\frac{\langle{\cal O}^{\cal H}[^3P_2^{(1)}]\rangle}{M^5}
F^{[^3P_2]}(t_1,t_2,\varphi),
\nonumber\\
\overline{|{\cal A}(R + R \to {\cal H}[^3S_1^{(8)}]|^2}
&=&\frac{1}{2}\pi^2\alpha_s^2\frac{\langle{\cal
O}^{\cal H}[^3S_1^{(8)}]\rangle}{M^3}
F^{[^3S_1]}(t_1,t_2,\varphi),
\nonumber\\
\overline{|{\cal A}(R + R \to {\cal H}[^1S_0^{(8)}]|^2}
&=&\frac{5}{12}\pi^2\alpha_s^2\frac{\langle{\cal
O}^{\cal H}[^1S_0^{(8)}]\rangle}{M^3}
F^{[^1S_0]}(t_1,t_2,\varphi),
\nonumber\\
\overline{|{\cal A}(R + R \to {\cal H}[^3P_0^{(8)}]|^2}
&=&5 \pi^2
\alpha_s^2\frac{\langle{\cal O}^{\cal H}[^3P_0^{(8)}]\rangle}{M^5}
F^{[^3P_0]}(t_1,t_2,\varphi),
\nonumber\\
\overline{|{\cal A}(R + R \to {\cal H}[^3P_1^{(8)}]|^2}
&=&10 \pi^2
\alpha_s^2\frac{\langle{\cal O}^{\cal H}[^3P_1^{(8)}]\rangle}{M^5}
F^{[^3P_1]}(t_1,t_2,\varphi),
\nonumber\\
\overline{|{\cal A}(R + R \to {\cal H}[^3P_2^{(8)}]|^2}
&=&\frac{4}{3}\pi^2 \alpha_s^2\frac{\langle{\cal
O}^{\cal H}[^3P_2^{(8)}]\rangle}{M^5}
F^{[^3P_2]}(t_1,t_2,\varphi),
\label{eq:vgvgJPsi3p28}
\end{eqnarray}
where
\begin{eqnarray}
F^{[^3S_1]}(t_1,t_2,\varphi)
&=&\frac{\left( M^2 + |{\bf p}_T|^{2}\right)
\left[ (t_1+t_2)^2 + M^2 \left(t_1+t_2-2\sqrt{t_1 t_2}
\cos\varphi\right)\right]}{(M^2+t_1+t_2)^2},
\nonumber\\
F^{[^1S_0]}(t_1,t_2,\varphi)
&=&2 \frac{M^2}{(M^2+t_1+t_2)^2}
\left(M^2+|{\bf p}_T|^{2}\right)^2\sin^2{\varphi},
\nonumber\\
F^{[^3P_0]}(t_1,t_2,\varphi)
&=&\frac{2}{9}\,\frac{M^2
\left(M^2 + |{\bf p}_T|^2 \right)^2\left[(3 M^2 +t_1+t_2)
\cos{\varphi} + 2\sqrt{t_1 t_2}\right]^2}{(M^2 + t_1 + t_2)^4},
\nonumber\\
F^{[^3P_1]}(t_1,t_2,\varphi)
&=&\frac{2}{9}\,\frac{M^2
\left(M^2 + |{\bf p}_T|^2 \right)^2
\left[(t_1+t_2)^2 \sin^2{\varphi} + M^2 \left(t_1+t_2-2\sqrt{t_1
t_2}\cos{\varphi}\right)\right]}{(M^2 + t_1 + t_2)^4},
\nonumber\\
F^{[^3P_2]}(t_1,t_2,\varphi)
&=&\frac{1}{3}\,
\frac{M^2}{(M^2+t_1+t_2)^4} \left(M^2+|{\bf p}_T|^2\right)^2 \left\{3 M^4 +
3M^2(t_1+t_2)+4t_1 t_2 \right.
\nonumber\\
&&{}+\left.
(t_1+t_2)^2\cos^2{\varphi}+2 \sqrt{t_1 t_2}\left[3
M^2 +2 (t_1+t_2)\right]\cos{\varphi}\right\}.
\end{eqnarray}
Here ${\bf p}_T={\bf k}_{1T}+{\bf k}_{2T}$,
$t_{1,2}= |{\bf k}_{{1,2}T}|^2$, and $\varphi=\varphi_1-\varphi_2$ is the
angle enclosed between ${\bf k}_{1T}$ and ${\bf k}_{2T}$, so that
\begin{equation}
|{\bf p}_T|^{2}=t_1+t_2+2\sqrt{t_1 t_2}\cos\varphi.
\end{equation}

It is interesting to consider the contribution of the diagram involving a
three-gluon vertex separately.
It is equal to
\begin{equation}
\overline{|{\cal A}_3(R + R \to {\cal
H}[^3S_1^{(8)}])|^2}=\pi^2\alpha_s^2\frac{\langle{\cal O}^{\cal
H}[^3S_1^{(8)}]\rangle}{2M^3} (M^2\cos ^2\varphi+|{\bf
p}_T|^{2}).
\end{equation}
For $|{\bf p}_T|^2 \gg M^2$, one has
\begin{equation}
\overline{|{\cal A}_3(R + R \to {\cal
H}[^3S_1^{(8)}])|^2}\approx\pi^2\alpha_s^2\frac{\langle{\cal
O}^{\cal H}[^3S_1^{(8)}]\rangle}{2M^3}|{\bf
p}_T|^{2},
\label{eq:vgvgJPsi3s18}
\end{equation}
which makes up the bulk of the contribution and can be interpreted as being
due to the fragmentation production of the ${\cal H}$ meson.
In fact, the right-hand side of Eq.~(\ref{eq:vgvgJPsi3s18}) can be written in
the factorized form
\begin{equation}
\overline{|{\cal A}_3(R + R \to {\cal H}[^3S_1^{(8)}])|^2}\approx
\overline{|{\cal A}(R + R \to g)|^2}P(g \to {\cal H}[^3S_1^{(8)}]),
\end{equation}
where
\begin{equation}
\overline{|{\cal A}(R + R \to g)|^2}=\frac{3}{2}\pi\alpha_s |{\bf p}_T|^{2}
\end{equation}
refers to real-gluon production by reggeized-gluon fusion \cite{PRD2003} and
\begin{equation}
P(g \to {\cal H}[^3S_1^{(8)}])=\pi\alpha_s
\frac{\langle{\cal O}^{\cal H}[^3S_1^{(8)}]\rangle}{3M^3}
\end{equation}
is the probability for the fragmentation of a gluon to a ${\cal H}$ meson,
which may be gleaned from the result for the corresponding fragmentation
function at the starting scale $\mu_0$ \cite{FFBraaten},
\begin{equation}
D_{g \to {\cal H}[^3S_1^{(8)}]}(z,\mu_0)
=\pi\alpha_s\frac{\langle{\cal O}^{\cal H}[^3S_1^{(8)}]\rangle}{3M^3}
\delta(1-z).
\end{equation}

The counterparts of Eq.~(\ref{eq:vgvgJPsi3p28}) in the collinear parton model
of QCD emerge through the operation
\begin{equation}
\overline{|{\cal A}(g + g \to {\cal
H}[^{2S+1}L_J^{(1,8)}]|^2}=\lim_{t_1,t_2\to0}\int_0^{2\pi}
\frac{d\varphi_1}{2\pi}\int_0^{2\pi}\frac{d\varphi_2}{2\pi}\overline{|{\cal
A}(R + R \to {\cal H}[^{2S+1}L_J^{(1,8)}]|^2}.
\label{eq:limit}
\end{equation}
In this way, we recover the well-known results \cite{Leibovich}:
\begin{eqnarray}
\overline{|{\cal A}(g + g \to {\cal H}[^{3}P_0^{(1)}]|^2}
&=&\frac{8}{3}\pi^2\alpha_s^2 \frac{\langle{\cal O}^{\cal
H}[^3P_0^{(1)}]\rangle}{M^3},
\nonumber\\
\overline{|{\cal A}(g + g \to {\cal H}[^{3}P_1^{(1)}]|^2}
&=&0,
\nonumber\\
\overline{|{\cal A}(g + g \to {\cal H}[^{3}P_2^{(1)}]|^2}
&=&\frac{32}{45}\pi^2\alpha_s^2 \frac{\langle{\cal O}^{\cal
H}[^3P_2^{(1)}]\rangle}{M^3},
\nonumber\\
\overline{|{\cal A}(g + g \to {\cal H}[^{3}S_1^{(8)}]|^2}
&=&0,
\nonumber\\
\overline{|{\cal A}(g + g \to {\cal H}[^{1}S_0^{(8)}]|^2}
&=&\frac{5}{12}\pi^2\alpha_s^2
\frac{\langle{\cal O}^{\cal H}[^1S_0^{(8)}]\rangle}{M},
\nonumber\\
\overline{|{\cal A}(g + g \to {\cal H}[^{3}P_0^{(8)}]|^2}
&=&5\pi^2\alpha_s^2\frac{\langle{\cal O}^{\cal H}[^3P_0^{(8)}]\rangle}{M^3},
\nonumber\\
\overline{|{\cal A}(g + g \to {\cal H}[^{3}P_1^{(8)}]|^2}
&=&0,
\nonumber\\
\overline{|{\cal A}(g + g \to {\cal H}[^{3}P_2^{(8)}]|^2}
&=&\frac{4}{3}\pi^2\alpha_s^2
\frac{\langle{\cal O}^{\cal H}[^3P_2^{(8)}]\rangle}{M^3}.
\end{eqnarray}

In the case of the $2\to2$ subprocess~(\ref{eq:RRtoHG}), we find
\begin{eqnarray}
\lefteqn{\overline{|{\cal A}(R + R \to {\cal H}[^{3}S_1^{(1)}] +
g|^2}=\pi^3 \alpha_s^3\frac{\langle{\cal O}^{\cal
H}[^3S_1^{(1)}]\rangle}{M^3}}
\nonumber\\
&&{}\times\frac{-320 M^4}{81
(M^2 - {\hat s})^2 (M^2 +t_1- {\hat t})^2 (M^2 +t_2- {\hat u})^2}
 (4 t_1 t_2 M^2 (t_1 + t_2 +
M^2)^2  \cos^4(\varphi_1 - \varphi_2)
\nonumber\\
&&{}- 2 \sqrt{t_1 t_2} (t_1 + t_2 + M^2) \cos^3(\varphi_1 -
\varphi_2) (-t_2^2 M^2 - 3 M^6 + t_2^2 {\hat t} + 3 M^4 {\hat t} +
t_2 M^2 {\hat u} + 3 M^4 {\hat u}
\nonumber\\
&&{}+ t_2 {\hat t} {\hat u} - M^2 {\hat t}
{\hat u} + t_1^2 (t_2 - M^2 + {\hat u}) + t_1 (t_2^2 + {\hat t}
(M^2 + {\hat u}) + t_2 (5 M^2 + {\hat t} + {\hat u}))
\nonumber\\
&&{}- 2 \sqrt{t_2} |{\bf p}_T| (t_1^2 - (t_2 - M^2) (M^2
- {\hat t}) + t_1 (t_2 + 2 M^2 + {\hat t})) \cos{\varphi_2})
\nonumber\\
&&{}- (-M^2 + {\hat t} + {\hat u})^2 (t_2^2 M^2 + t_2 M^4
+ t_1^2 (t_2 + M^2) + t_2 {\hat t}^2 + M^2 {\hat t}^2 + M^2 {\hat
u}^2
\nonumber\\
&&{}+ t_1 (t_2^2 + M^4 + {\hat u}^2 + 2 t_2 (4 M^2 +
{\hat t} + {\hat u})) - 2 \sqrt{t_2} |{\bf p}_T| (t_1^2 + (t_2 +
M^2) (M^2 - {\hat t})
\nonumber\\
&&{}+ t_1 (t_2 + 8 M^2 + {\hat t} +
2 {\hat u})) \cos{\varphi_2} + (t_1^2 + (t_2 + M^2)^2 + 2 t_1 (t_2
+ 5 M^2)) |{\bf p}_T|^2 \cos^2{\varphi_2})
\nonumber\\
&&{}+
\cos^2(\varphi_1 - \varphi_2) (t_2^3 M^4 - 2 t_2^2 M^6 + 5 t_2 M^8
- 2 t_2^3 M^2 {\hat t} + 4 t_2^2 M^4 {\hat t} - 10 t_2 M^6 {\hat
t} + t_2^3 {\hat t}^2
\nonumber\\
&&{}- 2 t_2^2 M^2 {\hat t}^2 + 5
t_2 M^4 {\hat t}^2 - t_2^3 M^2 {\hat u} - 7 t_2 M^6 {\hat u} +
t_2^3 {\hat t} {\hat u} - t_2^2 M^2 {\hat t} {\hat u} + 11 t_2 M^4
{\hat t} {\hat u} - 3 M^6 {\hat t} {\hat u}
\nonumber\\
&&{}+ t_2^2
{\hat t}^2 {\hat u} - 4 t_2 M^2 {\hat t}^2 {\hat u} + 3 M^4 {\hat
t}^2 {\hat u} + 2 t_2 M^4 {\hat u}^2 + t_2^2 {\hat t} {\hat u}^2 +
3 M^4 {\hat t} {\hat u}^2 - 2 M^2 {\hat t}^2 {\hat u}^2
\nonumber\\
&&{}+ t_1^3 (-4 t_2^2 + (M^2 - {\hat u}) (M^2 - {\hat t}
- {\hat u}) + t_2 (-5 M^2 + {\hat t} + {\hat u})) + t_1^2 (-4
t_2^3 - 2 M^6 + 4 M^4 {\hat u}
\nonumber\\
&&{}+ {\hat t} {\hat u}
({\hat t} + {\hat u}) - 2 t_2^2 (12 M^2 + {\hat t} + {\hat u}) -
M^2 {\hat u} ({\hat t} + 2 {\hat u}) + t_2 (-25 M^4 + {\hat t}^2 +
7 {\hat t} {\hat u} + 2 {\hat u}^2
\nonumber\\
&&{}+ M^2 (8 {\hat t}
+ 15 {\hat u}))) + t_1 (5 M^8 - 4 M^2 {\hat t} {\hat u}^2 + t_2^3
(-5 M^2 + {\hat t} + {\hat u}) - M^6 (7 {\hat t} + 10 {\hat u})
\nonumber\\
&&{}+ M^4 (2 {\hat t}^2 + 11 {\hat t} {\hat u} + 5 {\hat
u}^2) + t_2^2 (-25 M^4 + 2 {\hat t}^2 + 7 {\hat t} {\hat u} +
{\hat u}^2 + M^2 (15 {\hat t} + 8 {\hat u}))
\nonumber\\
&&{}+ t_2
(-11 M^6 + 15 M^4 ({\hat t} + {\hat u}) + 6 {\hat t} {\hat u}
({\hat t} + {\hat u}) - 2 M^2 (4 {\hat t}^2 + 3 {\hat t} {\hat u}
+ 4 {\hat u}^2)))
\nonumber\\
&&{}+ 2 \sqrt{t_2} |{\bf p}_T| (t_1^4 +
t_1^3 (3 t_2 + 11 M^2 + {\hat s} + {\hat t}) + t_1^2 (3 t_2^2 + 7
M^4 + {\hat s} {\hat t} + t_2 (33 M^2 - 2 {\hat s} + 3 {\hat t})
\nonumber\\
&&{}+ M^2 (11 {\hat s} + 4 {\hat t})) + t_1 (t_2^3 - 3
M^6 - 4 M^2 {\hat t}^2 + 7 M^4 ({\hat s} + {\hat t}) + t_2^2 (5
M^2 + {\hat s} + 3 {\hat t})
\nonumber\\
&&{}+ t_2 (27 M^4 + 2 M^2
({\hat s} - 10 {\hat t}) + 6 {\hat s} {\hat t})) - (M^2 - {\hat
t}) (t_2^3 - t_2 M^4 + t_2^2 {\hat s}
\nonumber\\
&&{}+ M^2 (-4 {\hat
t} ({\hat s} + {\hat t}) + M^2 (3 {\hat s} + 4 {\hat t}))))
\cos{\varphi_2}\ - 4 t_2 |{\bf p}_T|^2 (t_1^3 + 2 t_1^2 (t_2 + 2
M^2) + 2 (M^3 - M {\hat t})^2
\nonumber\\
&&{}+ t_1 (t_2^2 + 2 t_2
M^2 + 5 M^4 - 4 M^2 {\hat t})) \cos^2{\varphi_2}) - |{\bf p}_T|^2
\cos^2{\varphi_1} (2 t_1^3 t_2 + 4 t_1^2 t_2^2 + 2 t_1 t_2^3
\nonumber\\
&&{}+ 4 t_1^2 t_2 M^2 + 8 t_1 t_2^2 M^2 + t_1^2 M^4 + 12
t_1 t_2 M^4 + t_2^2 M^4 + 6 t_1 M^6 + 10 t_2 M^6 + M^8
\nonumber\\
&&{}+ 4 M^6 |{\bf p}_T|^2 - 2 t_1^2 M^2 {\hat t} - 4 t_1
t_2 M^2 {\hat t} - 2 t_2^2 M^2 {\hat t} - 4 t_1 M^4 {\hat t} - 20
t_2 M^4 {\hat t} - 2 M^6 {\hat t}
\nonumber\\
&&{}- 8 M^4 |{\bf
p}_T|^2 {\hat t} + t_1^2 {\hat t}^2 + 2 t_1 t_2 {\hat t}^2 + t_2^2
{\hat t}^2 + 2 t_1 M^2 {\hat t}^2 + 10 t_2 M^2 {\hat t}^2 + M^4
{\hat t}^2 + 4 M^2 |{\bf p}_T|^2 {\hat t}^2
\nonumber\\
&&{}- 2 t_1^2
M^2 {\hat u} - 12 t_1 t_2 M^2 {\hat u} - 2 t_2^2 M^2 {\hat u} - 12
t_1 M^4 {\hat u} - 20 t_2 M^4 {\hat u} - 2 M^6 {\hat u} - 8 M^4
|{\bf p}_T|^2 {\hat u}
\nonumber\\
&&{}+ 2 t_1^2 {\hat t} {\hat u} +
4 t_1 t_2 {\hat t} {\hat u} + 2 t_2^2 {\hat t} {\hat u} + 4 t_1
M^2 {\hat t} {\hat u} + 20 t_2 M^2 {\hat t} {\hat u} + 2 M^4 {\hat
t} {\hat u} + 8 M^2 |{\bf p}_T|^2 {\hat t} {\hat u} + t_1^2 {\hat
u}^2
\nonumber\\
&&{}+ 2 t_1 t_2 {\hat u}^2 + t_2^2 {\hat u}^2 + 6
t_1 M^2 {\hat u}^2 + 10 t_2 M^2 {\hat u}^2 + M^4 {\hat u}^2 + 4
M^2 |{\bf p}_T|^2 {\hat u}^2
\nonumber\\
&&{}- 8 \sqrt{t_1} M^2 |{\bf
p}_T| (t_2 + M^2 - {\hat u}) (M^2 - {\hat t} - {\hat u})
\Bigr(\cos{\varphi_1} + \cos(\varphi_1 - 2
\varphi_2)\Bigr)
\nonumber\\
&&{} + 4 \sqrt{t_1 t_2} \cos(\varphi_1 -
\varphi_2) \Bigl(M^2-{\hat t}-{\hat u}\Bigr) \Bigl((t_1 + t_2)^2 +
M^2 (5 M^2 + 2 t_1 + 6 t_2 - 4 {\hat u})\Bigr)
\nonumber\\
&&{} + 2
t_1 \cos(2 (\varphi_1 - \varphi_2)) \Bigl( t_2 \Bigl( (t_1 +
t_2)^2 + M^2 (2 t_1 + 4 t_2 + 5 M^2 - 4 {\hat u}) \Bigr) + 2 M^2
\Bigl( M^2 - {\hat u} \Bigr)^2\Bigr)
\nonumber\\
&&{}- 16 M^2 |{\bf
p}_T| \sqrt{t_2} \cos{\varphi_2} \Bigr(M^2 - {\hat t} - {\hat u}
\Bigl)^2 + 4 M^2 |{\bf p}_T|^2 \cos(2 \varphi_2) \Bigr( M^2 (M^2 -
2 {\hat t} - 2 {\hat u}) + ({\hat t} + {\hat u})^2 \Bigl)
\nonumber\\
&&{}- 2 \sqrt{t_1} (M^2 - {\hat t} - {\hat u})
\cos(\varphi_1 - \varphi_2) (-2 |{\bf p}_T| (2 t_1^2 t_2 - M^6 + 2
t_2^2 (M^2 - {\hat t}) + M^4 {\hat t} + M^4 {\hat u} - 2 M^2 {\hat
t} {\hat u}
\nonumber\\
&&{}+ t_2 (8 M^4 - 7 M^2 {\hat t} - {\hat t}
({\hat t} + {\hat u})) + t_1 (2 t_2^2 + M^2 (3 M^2 - {\hat t} - 3
{\hat u}) + t_2 (11 M^2 + {\hat t} + 3 {\hat u}))) \cos{\varphi_2}
\nonumber\\
&&{}+ \sqrt{t_2} (2 t_2^2 M^2 + 7 t_2 M^4 - M^6 + t_2^2
|{\bf p}_T|^2 + 2 t_2 M^2 |{\bf p}_T|^2 + 5 M^4 |{\bf p}_T|^2
\nonumber\\
&&{}+ t_1^2 (2 t_2 + 2 M^2 + |{\bf p}_T|^2) - 4 t_2 M^2
{\hat t} - 4 M^2 |{\bf p}_T|^2 {\hat t} + t_2 {\hat t}^2 + 3 M^2
{\hat t}^2 - t_2 M^2 {\hat u} - t_2 {\hat t} {\hat u}
\nonumber\\
&&{}- M^2 {\hat t} {\hat u} - {\hat t}^2 {\hat u} + 3 M^2
{\hat u}^2 - {\hat t} {\hat u}^2 + t_1 (2 t_2^2 + 7 M^4 + M^2 (6
|{\bf p}_T|^2 - {\hat t} - 4 {\hat u}) + {\hat u} (-{\hat t} +
{\hat u})
\nonumber\\
&&{}+ t_2 (13 M^2 + 2 |{\bf p}_T|^2 + 3 ({\hat
t} + {\hat u}))) + |{\bf p}_T|^2 (t_1^2 + t_2^2 + 2 t_2 M^2 + 5
M^4 + 2 t_1 (t_2 + 3 M^2)
\nonumber\\
&&{}- 4 M^2 {\hat t}) \cos(2
\varphi_2))) + 2 |{\bf p}_T| \cos{\varphi_1} (2 t_1 \sqrt{t_2}
(t_1 + t_2 + M^2) (t_2^2 + M^4 - M^2 {\hat u}
\nonumber\\
&&{}+ t_1
(t_2 - M^2 + {\hat u}) + t_2 (2 M^2 + {\hat u})) \cos^3(\varphi_1
- \varphi_2) + \sqrt{t_1} \cos^2(\varphi_1 - \varphi_2) (t_2^4 +
11 t_2^3 M^2 + 7 t_2^2 M^4
\nonumber\\
&&{}- 3 t_2 M^6 + t_2^3 {\hat
s} + 11 t_2^2 M^2 {\hat s} + 7 t_2 M^4 {\hat s} - 3 M^6 {\hat s} +
t_2^3 {\hat u} + 4 t_2^2 M^2 {\hat u} + 7 t_2 M^4 {\hat u} - 4 M^6
{\hat u}
\nonumber\\
&&{}+ t_2^2 {\hat s} {\hat u} + 7 M^4 {\hat s}
{\hat u} - 4 t_2 M^2 {\hat u}^2 + 8 M^4 {\hat u}^2 - 4 M^2 {\hat
s} {\hat u}^2 - 4 M^2 {\hat u}^3 + t_1^3 (t_2 - M^2 + {\hat u})
\nonumber\\
&&{}+ t_1^2 (3 t_2^2 + {\hat s} (-M^2 + {\hat u}) + t_2
(5 M^2 + {\hat s} + 3 {\hat u})) + t_1 (3 t_2^3 + M^6 - M^4 {\hat
u}
\nonumber\\
&&{}+ t_2^2 (33 M^2 - 2 {\hat s} + 3 {\hat u}) + t_2
(27 M^4 + 2 M^2 ({\hat s} - 10 {\hat u}) + 6 {\hat s} {\hat u})) +
2 \sqrt{t_2} |{\bf p}_T| (-5 M^6 + 5 M^4 {\hat t}
\nonumber\\
&&{}- 4
t_2 (M^4 - M^2 {\hat t}) + t_1^2 (M^2 - {\hat t} - {\hat u}) +
t_2^2 (M^2 - {\hat t} - {\hat u}) + 5 M^4 {\hat u} - 2 t_1 (2 M^2
(M^2 - {\hat u})
\nonumber\\
&&{}+ t_2 (M^2 + {\hat t} + {\hat u})))
\cos{\varphi_2}) + \sqrt{t_1} (-M^2 + {\hat t} + {\hat u})^2 (t_1
t_2 + t_2^2 + t_1 M^2 + 8 t_2 M^2 + M^4
\nonumber\\
&&{}+ 4 M^2 |{\bf
p}_T|^2 + 2 t_2 {\hat t} - t_1 {\hat u} + t_2 {\hat u} - M^2 {\hat
u} - 2 \sqrt{t_2} |{\bf p}_T| (7 M^2 + {\hat t} + {\hat u})
\cos{\varphi_2} + 4 M^2 |{\bf p}_T|^2 \cos(2 \varphi_2))
\nonumber\\
&&{}- (M^2 - {\hat t} - {\hat u}) \cos(\varphi_1 -
\varphi_2) (-(|{\bf p}_T| (3 M^6 - 3 M^4 {\hat t} - 12 t_2 (M^4 -
M^2 {\hat t}) + t_1^2 (M^2 - {\hat t} - {\hat u})
\nonumber\\
&&{}+
t_2^2 (M^2 - {\hat t} - {\hat u}) - 3 M^4 {\hat u} + 4 M^2 {\hat
t} {\hat u} - 2 t_1 (6 M^2 (M^2 - {\hat u}) + t_2 (11 M^2 + 3
({\hat t} + {\hat u})))) \cos{\varphi_2})
\nonumber\\
&&{}- 2
\sqrt{t_2} (2 t_1^2 (t_2 + M^2 - {\hat u}) - M^2 (M^4 + 2 {\hat t}
(|{\bf p}_T|^2 + {\hat u}) - M^2 (2 |{\bf p}_T|^2 + {\hat t} +
{\hat u})
\nonumber\\
&&{}+ t_2 (-3 M^2 + 3 {\hat t} + {\hat u})) +
t_1 (2 t_2^2 + 8 M^4 + M^2 (2 |{\bf p}_T|^2 - 7 {\hat u}) - {\hat
u} ({\hat t} + {\hat u})
\nonumber\\
&&{}+ t_2 (11 M^2 + 3 {\hat t} +
{\hat u})) + 2 M^2 |{\bf p}_T|^2 (t_1 + M^2 - {\hat t}) \cos(2
\varphi_2))))),
\label{eq:RRtoeHG}
\end{eqnarray}
where $\hat s=(k_1+k_2)^2$, $\hat t=(k_1-p)^2$, and $\hat u=(k_2-p)^2$ are the
standard Mandelstam variables.
With the aid of Eq.~(\ref{eq:limit}), we recover from Eq.~(\ref{eq:RRtoeHG})
the well-known collinear-parton-model result \cite{Leibovich},
\begin{eqnarray}
\overline{|{\cal A}(g + g \to {\cal H}[^{3}S_1^{(1)}] + g|^2}
&=&\pi^3 \alpha_s^3 \frac{\langle{\cal O}^{\cal
H}[^3S_1^{(1)}]\rangle}{M^3}\, \frac{320 M^4}{81 (M^2 - {\hat t})^2
(M^2 - {\hat u})^2 ({\hat t} + {\hat u})^2}
\nonumber\\
&&{}\times (M^4
{\hat t}^2 - 2 M^2 {\hat t}^3 + {\hat t}^4 + M^4 {\hat t} {\hat u}
- 3 M^2 {\hat t}^2 {\hat u} + 2 {\hat t}^3 {\hat u} + M^4 {\hat
u}^2
\nonumber\\
&&{} - 3 M^2 {\hat t} {\hat u}^2 + 3 {\hat t}^2
{\hat u}^2 - 2 M^2 {\hat u}^3 + 2 {\hat t} {\hat u}^3 + {\hat
u}^4).
\end{eqnarray}

We now turn to subprocesses~(\ref{eq:FRtoH}) and (\ref{eq:FRtoHG}), with one
real photon in the initial state.
For the $2 \to 1$ subprocesses (\ref{eq:FRtoH}), which are pure color-octet
processes, we find
\begin{eqnarray}
\overline{|{\cal A}(R + \gamma \to {\cal
H}[^{3}S_1^{(8)}]|^2}&=&0,
\nonumber\\
\overline{|{\cal A}(R + \gamma \to {\cal H}[^{1}S_0^{(8)}]|^2}&=&8
\pi^2 \alpha \alpha_s e_Q^2 \frac{\langle{\cal O}^{\cal
H}[^1S_0^{(8)}]\rangle}{M},
\nonumber\\
\overline{|{\cal A}(R + \gamma \to {\cal
H}[^{3}P_0^{(8)}]|^2}&=&\frac{32}{3} \pi^2\alpha \alpha_s
e_Q^2\frac{\langle{\cal O}^{\cal
H}[^3P_0^{(8)}]\rangle}{M^3}\,\frac{(3 M^2 + t_1)^2}{(M^2 +
t_1)^2},
\nonumber\\
\overline{|{\cal A}(R + \gamma \to {\cal
H}[^{3}P_1^{(8)}]|^2}&=&\frac{64}{3} \pi^2\alpha \alpha_s
e_Q^2\frac{\langle{\cal O}^{\cal
H}[^3P_1^{(8)}]\rangle}{M^3}\,\frac{t_1(2 M^2 + t_1)}{(M^2 +
t_1)^2},
\nonumber\\
\overline{|{\cal A}(R + \gamma \to {\cal
H}[^{3}P_2^{(8)}]|^2}&=&\frac{64}{15}\pi^2\alpha \alpha_s
e_Q^2\frac{\langle{\cal O}^{\cal
H}[^3P_2^{(8)}]\rangle}{M^3}\,\frac{6 M^4 + 6 M^2 t_1 +
t_1^2}{(M^2 + t_1)^2}, \label{eq:gammaR}
\end{eqnarray}
where $e_Q$ is electric charge of the heavy quark $Q$.
Application of Eq.~(\ref{eq:limit}) to Eq.~(\ref{eq:gammaR}) yields the
well-known results of the collinear parton model \cite{KraemerGammaP},
\begin{eqnarray}
\overline{|{\cal A}(g +\gamma \to {\cal H}[^{3}S_1^{(8)}]|^2}&=&0,
\nonumber\\
\overline{|{\cal A}(g +\gamma \to {\cal H}[^{1}S_0^{(8)}]|^2}&=&8
\pi^2 \alpha \alpha_s e_Q^2\frac{\langle{\cal O}^{\cal
H}[^1S_0^{(8)}]\rangle}{M},
\nonumber\\
\overline{|{\cal A}(g +\gamma \to {\cal H}[^{3}P_0^{(8)}]|^2}&=&
96 \pi^2\alpha \alpha_s e_Q^2\frac{\langle{\cal O}^{\cal
H}[^3P_0^{(8)}]\rangle}{M^3},
\nonumber\\
\overline{|{\cal A}(g +\gamma \to {\cal H}[^{3}P_1^{(8)}]|^2}&=&
0,
\nonumber\\
\overline{|{\cal A}(g +\gamma \to {\cal
H}[^{3}P_2^{(8)}]|^2}&=&\frac{128}{5}\pi^2\alpha \alpha_s
e_Q^2\frac{\langle{\cal O}^{\cal H}[^3P_2^{(8)}]\rangle}{M^3}.
\end{eqnarray}

For the $2 \to 2$ subprocess (\ref{eq:FRtoHG}), which is a color-singlet
process, we find
\begin{eqnarray}
\lefteqn{\overline{|{\cal A}(R + \gamma \to {\cal
H}[^{3}S_1^{(1)}] + g|^2}=\pi^3\alpha \alpha_s^2
e_Q^2\frac{\langle{\cal O}^{\cal H}[^3S_1^{(1)}]\rangle}{M^3}\,
\frac{2048 M^2}{27 (M^2 - {\hat s})^2 (M^2 - {\hat u})^2 (t_1 +
M^2 - {\hat t})^2}}
\nonumber\\
&&{}\times \Bigl( t_1^4 M^2 + M^2 \bigl({\hat s}^2 + {\hat s}
{\hat u} + {\hat u}^2 - M^2 ({\hat s} + {\hat u})\bigr)^2 + t_1^3
(M^2 (5 {\hat s} + 3 {\hat u})-7 M^4 - {\hat s} {\hat u})
\nonumber\\
&&{}+ t_1^2 ({\hat s} {\hat u} ({\hat u}-{\hat s})+ M^4 (3 {\hat
u}-11 {\hat s}) + M^2 (7 {\hat s}^2 + 2 {\hat s} {\hat u} - 3
{\hat u}^2)) + t_1 {\hat s} ({\hat s} {\hat u}^2 + M^4 ({\hat
u}-6 {\hat s})
\nonumber\\
&&{} + M^2 (4 {\hat s}^2 + {\hat s} {\hat u} - {\hat u}^2))- 2
\sqrt{t_1} |{\bf p}_T| (t_1^3 M^2 + t_1^2 (-7 M^4 - {\hat s}
{\hat u} + M^2 (3 {\hat s} + 4 {\hat u}))
\nonumber\\
&&{} + t_1 (M^4 (-7 {\hat s} + 2 {\hat u}) -{\hat s}^2 {\hat u}+
M^2 (2 {\hat s}^2 + {\hat s} {\hat u} - 2 {\hat u}^2)) - M^2 (2
M^4 ({\hat s} + {\hat u}) - 2 M^2 {\hat u} (3 {\hat s} + 2 {\hat
u}) \nonumber\\ &&{} + {\hat u} (3 {\hat s}^2 + 4 {\hat s} {\hat
u} + 2 {\hat u}^2))) \cos{\varphi} - 2 M^2 |{\bf p}_T|^2 (t_1^3 +
M^2 {\hat s}^2 + t_1^2 (M^2 + 2 {\hat s}) \nonumber\\
&&{}+ t_1 (2 M^2 {\hat s} + {\hat s}^2 - 2 {\hat t}^2))
\cos^2{\varphi}\Bigr), \label{eq:gammaRg}
\end{eqnarray}
where $k_2^\mu$ now represents the photon four-momentum and
$\varphi_2=0$.
Equation~(\ref{eq:gammaRg}) agrees with the
corresponding result in Ref.~\cite{ZotovLEP}, but has a more
compact form. By means of Eq.~(\ref{eq:limit}),
Eq.~(\ref{eq:gammaRg}) collapses to the well-known
collinear-parton model result \cite{kls},
\begin{eqnarray}
\overline{|{\cal A}(g + \gamma \to {\cal H}[^{3}S_1^{(1)}] +
g|^2}&=&\pi^3\alpha \alpha_s^2 e_Q^2\frac{\langle{\cal O}^{\cal
H}[^3S_1^{(1)}]\rangle}{M^3}\, \frac{2048 M^4}{27 (M^2 - {\hat
t})^2 (M^2 - {\hat u})^2 ({\hat t} + {\hat u})^2}
\nonumber\\
&&{}\times (M^4
{\hat t}^2 - 2 M^2 {\hat t}^3 + {\hat t}^4 + M^4 {\hat t} {\hat u}
- 3 M^2 {\hat t}^2 {\hat u} + 2 {\hat t}^3 {\hat u} + M^4 {\hat
u}^2
\nonumber\\
&&{} - 3 M^2 {\hat t} {\hat u}^2 + 3 {\hat t}^2
{\hat u}^2 - 2 M^2 {\hat u}^3 + 2 {\hat t} {\hat u}^3 + {\hat
u}^4).
\end{eqnarray}

Finally, we turn to subprocesses (\ref{eq:eRtoH}) and (\ref{eq:eRtoHG}),
through which electroproduction proceeds at LO.
As for the $2\to2$ subprocesses (\ref{eq:eRtoH}), which are all color-octet
processes, we have
\begin{eqnarray}
\lefteqn{\overline{|{\cal A}(R + e \to e + {\cal H}[^{3}S_1^{(8)}]|^2}=0,}
\nonumber\\
\lefteqn{\overline{|{\cal A}(R + e \to e + {\cal
H}[^{1}S_0^{(8)}]|^2}=64 \pi^3\alpha^2 \alpha_s
e_Q^2\frac{\langle{\cal O}^{\cal H}[^1S_0^{(8)}]\rangle}{M^3}\,
\frac{1}{y_2^2 Q^2 (M^2 + Q^2 + t_1)^2}}
\nonumber\\
&&{}\times\biggl( (2 + (y_2 - 2) y_2) \Bigl((M^2 + t_1)^2 + Q^4 +
2 Q^2 M^2 - 2 Q^2 t_1 y_2\Bigr) +
    4 Q^2 t_1 \nonumber\\&&+
    4 Q \sqrt{t_1 (1 - y_2)} (M^2 + Q^2 + t_1) (y_2 - 2) y_2 \cos{(\varphi_1 -
 \varphi_2)}
\nonumber\\
&&{}+ 2 (y_2 - 1) \Bigl(Q^4 + (M^2 + t_1)^2 + Q^2 (2 M^2 + 2 t_1 -
t_1 y_2^2)\Bigr) \cos{(2 (\varphi_1 - \varphi_2))}\biggr) M^2,
\nonumber\\
\lefteqn{\overline{|{\cal A}(R + e \to e + {\cal
H}[^{3}P_0^{(8)}]|^2}=\frac{256}{3} \pi^3\alpha^2 \alpha_s
e_Q^2\frac{\langle{\cal O}^{\cal
H}[^3P_0^{(8)}]\rangle}{M^5}\,\frac{1}{y_2^2 Q^2 (M^2 + Q^2 +
t_1)^4}}
\nonumber\\
&&{}\times\biggl((2 + (y_2 - 2) y_2)\Bigl(9 M^8 + 24 M^6 (Q^2 +
t_1) + 22 M^4 Q^4 + 22 M^4 t_1^2 +
    (Q^2 + t_1)^2 (Q^4 + t_1^2)
\nonumber\\
&&{}+ 8 M^2 (Q^2 + t_1) (Q^4 + t_1^2)\Bigr) +
    2 M^4 Q^2 t_1 (52 + y_2 ((43 - 9 y_2) y_2 -64))
\nonumber\\
&&{}+
    2 Q^2 t_1 (Q^2 + t_1)^2 (10 - y_2 (14 + (y_2 - 6) y_2)) \nonumber\\
&&{}+
    4 M^2 Q^2 t_1 (Q^2 + t_1) (16 - 3 y_2 (8 + (y_2 - 5) y_2))
\nonumber\\
&&{}+
    4 Q \sqrt{t_1 (1 - y_2)} (M^2 + Q^2 + t_1)
     (Q^4 (4 + (y_2 - 2) y_2) + 2 Q^2 (t_1 (4 + (y_2 - 6) y_2)
\nonumber\\
&&{}+ M^2 (8 + y_2 (-4 + 3 y_2))) +
       (3 M^2 + t_1) (t_1 (4 + (y_2 - 2) y_2)
\nonumber\\
&&{}+ M^2 (4 + y_2 (3 y_2 - 2))))
 \cos{(\varphi_1 - \varphi_2)}
-
    2 (3 M^2 + Q^2 + t_1) (y_2 - 1)\nonumber\\
&&{}\times
     (3 M^6 + 7 M^4 (Q^2 + t_1) + (Q^2 + t_1) (Q^4 + t_1^2 + Q^2 t_1 (2 +
 (y_2 - 4) y_2))
\nonumber\\
&&{}+
       M^2 (5 Q^4 + 5 t_1^2 + Q^2 t_1 (10 + y_2 (3 y_2-4)))) \cos{(2 (\varphi_1 -
       \varphi_2))}\biggr) M^2,
\nonumber\\
\lefteqn{\overline{|{\cal A}(R + e \to e + {\cal
H}[^{3}P_1^{(8)}]|^2}=\frac{512}{3} \pi^3\alpha^2 \alpha_s
e_Q^2\frac{\langle{\cal O}^{\cal H}[^3P_1^{(8)}]\rangle}{M^5}\,
\frac{1}{y_2^2 Q^2 (M^2 + Q^2 + t_1)^4 }}
\nonumber\\
&&{}\times\biggl((2 + (y_2 - 2) y_2)\Bigl(Q^8 + t_1 (M^2 + t_1)^2
(2 M^2 + t_1) \Bigr) \nonumber\\
&&{}+ 2 Q^6 (y_2 - 2) (M^2 (y_2 - 2) - t_1 (2 - y_2 + y_2^2))
\nonumber\\
&&{}+
    Q^4 (4 M^2 t_1 (y_2 - 3) (y_2 - 2) + M^4 (10 + (y_2-10) y_2) -
       2 t_1^2 (y_2 (6 + y_2 (-5 + 2 y_2))-6))
\nonumber\\
&&{}+
    2 Q^2 (M^4 t_1 (10 + (-8 + y_2) y_2) -2 M^6 (y_2 - 1) - t_1^3 (y_2 - 2) (2 +
 (y_2 - 1) y_2)
\nonumber\\
&&{}-
       M^2 t_1^2 (y_2 (10 + y_2 (2 y_2-5))-12)) +
    4 Q \sqrt{t_1 (1 - y_2)} (M^2 + Q^2 + t_1)
     (M^4 (y_2 - 2)
\nonumber\\
&&{}+ (Q^2 + t_1)^2 (y_2 - 2) y_2 - M^2 (Q^2 (2 + y_2) + t_1 (2 +
y_2 - 2 y_2^2)))
     \cos{(\varphi_1 - \varphi_2)} \nonumber\\
&&{}+ 2 (y_2 - 1) ((Q^2 + t_1)^4 - Q^2 t_1 (Q^2 + t_1)^2 y_2^2 +
       M^4 ((Q^2 + t_1)^2 - 2 Q^2 t_1 y_2) +
       2 M^2 ((Q^2 + t_1)^3
\nonumber\\
&&{}- Q^2 t_1 (Q^2 + t_1) y_2 - Q^2 t_1^2 y_2^2)) \cos{(2
(\varphi_1 -
       \varphi_2))}\biggr) M^2,
\nonumber\\
\lefteqn{ \overline{|{\cal A}(R + e \to e + {\cal
H}[^{3}P_2^{(8)}]|^2}=\frac{512}{15} \pi^3\alpha^2 \alpha_s
e_Q^2\frac{\langle{\cal O}^{\cal H}[^3P_2^{(8)}]\rangle}{M^5}\,
\frac{1}{y_2^2 Q^2 (M^2 + Q^2 + t_1)^4 }}
\nonumber\\
&&{}\times\biggl((2 + (y_2 - 2) y_2) \Bigl( Q^8+
    (M^2 + t_1)^2 (6 M^4 + 6 M^2 t_1 + t_1^2) \Bigr) +
    2 Q^6 (M^2 (8 + (y_2-8) y_2)
\nonumber\\
&&{}-t_1 (y_2 - 3) (y_2 - 2)^2) +
    Q^4 (M^4 (38 + y_2 (7 y_2-38)) + 4 M^2 t_1 (20 + y_2 (8 y_2-25))
\nonumber\\
&&{}+
       t_1^2 (44 - 2 y_2 (30 + y_2 (2 y_2-13)))) +
    2 Q^2 (-(t_1^3 (y_2 - 3) (y_2 - 2)^2) + 6 M^6 (3 + (y_2 - 3) y_2)
\nonumber\\
&&{}-
       M^2 t_1^2 (y_2 (50 + y_2 (6 y_2-25))-40) - M^4 t_1 (y_2 (52 + y_2 (6 y_2-25))-46))
\nonumber\\
&&{}+
    4 Q \sqrt{t_1 (1 - y_2)} (M^2 + Q^2 + t_1)(Q^4 (4 + (y_2 - 2) y_2) \nonumber\\
&&{}+ t_1^2 (4 +
 (y_2 - 2) y_2) + 3 M^4 (2 + y_2 (2 y_2-3))
+
       M^2 t_1 (10 + y_2 (6 y_2-11)) \nonumber\\
&&{}+ Q^2 (M^2 (10 - 11 y_2) + 2 t_1 (4 +
 (y_2 - 6) y_2)))
     \cos{(\varphi_1 - \varphi_2)}
\nonumber\\
&&{}- 2 (y_2 - 1) (2 M^2
        ((Q^2 + t_1)^3 - 5 Q^2 t_1 (Q^2 + t_1) y_2 + 3 Q^2 t_1^2 y_2^2) \nonumber\\
&&{}+
       (Q^2 + t_1)^2 (Q^4 + t_1^2
+ Q^2 t_1 (2 + (y_2 - 4) y_2)) \nonumber\\
&&{}+
       M^4 (Q^4 + t_1^2+ 2 Q^2 t_1 (1 + 3 (y_2 - 1) y_2))) \cos{(2 (\varphi_1 -
       \varphi_2))}\biggr) M^2.
\end{eqnarray}
As usual, $Q^2=-q^2$ and $y_2=(q\cdot P)/(k\cdot P)$, where $P^\mu$,
$k^\mu$, $k^{\prime\mu}$, and $q^\mu=k^\mu-k^{\prime\mu}$ are the four-momenta
of the incoming proton, the incoming lepton, the outgoing lepton, and the
virtual photon, respectively, $\varphi_1$ is the angle between
${\bf k}_{1T}$ and ${\bf p}_{T}$, and $\varphi_2$ is the angle between
${\bf q}_{T}$ and ${\bf p}_{T}$. The corresponding formulas in the
collinear parton model \cite{fm} are recovered as explained in
Eq.~(\ref{eq:limit}) and read:
\begin{eqnarray}
\overline{|{\cal A}(g + e \to e + {\cal H}[^{3}S_1^{(8)}]|^2}&=&0,
\nonumber\\
\overline{|{\cal A}(g + e \to e + {\cal H}[^{1}S_0^{(8)}]|^2}&=&64
\pi^3\alpha^2 \alpha_s e_Q^2\frac{\langle{\cal O}^{\cal
H}[^1S_0^{(8)}]\rangle}{M}\,\frac{y_2^2 - 2 y_2 + 2}{y_2^2 Q^2},
\nonumber\\
\overline{|{\cal A}(g + e \to e + {\cal
H}[^{3}P_0^{(8)}]|^2}&=&\frac{256}{3} \pi^3\alpha^2 \alpha_s
e_Q^2\frac{\langle{\cal O}^{\cal
H}[^3P_0^{(8)}]\rangle}{M^3}
\nonumber\\
&&{}\times\frac{(y_2^2 - 2 y_2 + 2) (Q^2 + 3 M^2)^2}{y_2^2 Q^2
(Q^2 + M^2)^2},
\nonumber\\
\overline{|{\cal A}(g + e \to e + {\cal
H}[^{3}P_1^{(8)}]|^2}&=&\frac{512}{3} \pi^3\alpha^2 \alpha_s
e_Q^2\frac{\langle{\cal O}^{\cal
H}[^3P_1^{(8)}]\rangle}{M^3}
\nonumber\\
&&{}\times\frac{((y_2^2 - 2 y_2 + 2)Q^2 -4 (y_2 - 1)) M^2}{y_2^2
(Q^2+M^2)^2},
\nonumber\\
\overline{|{\cal A}(g + e \to e + {\cal
H}[^{3}P_2^{(8)}]|^2}&=&\frac{512}{15} \pi^3\alpha^2 \alpha_s
e_Q^2\frac{\langle{\cal O}^{\cal
H}[^3P_2^{(8)}]\rangle}{M^3}
\nonumber\\
&&{}\times\frac{((y_2^2 - 2 y_2 + 2)(Q^4 + 6 M^4) -12 (y_2 - 1)
M^2 Q^2)}{y_2^2 Q^2 (Q^2+M^2)^2}.
\end{eqnarray}

Our analytic result for the $2\to3$ color-singlet subprocess (\ref{eq:eRtoHG})
is rather lengthy, and we refrain from listing it here.

\section{\label{sec:five}Charmonium production at the Tevatron}

During the last decade, the CDF Collaboration at the Tevatron
\cite{CDFI,CDFII} collected data on charmonium production at energies
$\sqrt{S}=1.8$~TeV (run~I) and $\sqrt{S}=1.96$~TeV (run~II) in the central
region of pseudorapidity $|\eta|<0.6$.
The data cover a large interval in transverse momentum, namely
$5<p_T<20$~GeV (run I) and $0<p_T<20$~GeV (run II).
The data sample of run~I \cite{CDFI} includes $p_T$ distributions of
$J/\psi$ mesons that were produced directly in the hard interaction, via
radiative decays of $\chi_{cJ}$ mesons, via decays of $\psi^\prime$ mesons,
and via decays of $b$ hadrons.
That of run~II \cite{CDFII} includes $p_T$ distributions of prompt
$J/\psi$ mesons, so far without separation into direct, $\chi_{cJ}$-decay, and
$\psi^\prime$-decay contributions, and of $J/\psi$ mesons from $b$-hadron
decays.

As is well known, the cross section of charmonium production measured at the
Tevatron is more than one order of magnitude larger than the prediction of the
CSM evaluated within the collinear parton model \cite{KramerReview}.
Switching from the collinear parton model to the $k_T$-factorization approach
\cite{CDFBaranov,KTTeryaev,KTYuan} somewhat ameliorates the situation, but
still does not lead to agreement at all.
On the other hand, a successful description of the data could be achieved with
the NRQCD factorization formalism \cite{NRQCD} implemented in the collinear
parton model, including the fusion and fragmentation mechanisms of charmonium
hadroproduction \cite{PMBraaten,BKLee}.

Charmonium hadroproduction was studied some time ago using the NRQCD
factorization formalism implemented in the $k_T$-factorization approach
invoking both the fusion \cite{CDFBaranov,KTTeryaev,KTYuan} and fragmentation
pictures \cite{PRD2003}.
It was found \cite{CDFBaranov,KTTeryaev,KTYuan} that, in order to describe the
experimental data from the CDF Collaboration \cite{CDFI}, it is necessary to
employ a set of NMEs that greatly differs from the one favored by the
collinear parton model.
In this paper, we confirm this conclusion only to some degree.

On the other hand, the polarization of prompt $J/\psi$ mesons measured at the
Tevatron \cite{CDFPolarization} also provides a sensitive probe of the NRQCD
mechanism.
This issue was carefully investigated both in the collinear parton model
\cite{PMPolarization} and in the $k_T$-factorization approach
\cite{KTPolarization}.
None of these studies was able to prove or disprove the NRQCD factorization
hypothesis.

In contrast to previous analyses in the collinear parton model or
the $k_T$-factorization approach, we perform a joint fit to the
run-I and run-II CDF data \cite{CDFI,CDFII} to obtain the
color-octet NMEs for $J/\psi$, $\psi^\prime$, and $\chi_{cJ}$
mesons. We use three different versions of unintegrated gluon
distribution function. Our calculations are based on exact
analytical expressions for the relevant squared amplitudes, which
were previously unknown in literature. Our fits include five
experimental data sets, which come as $p_T$ distributions of
$J/\psi$ mesons from direct production, prompt production,
$\chi_{cJ}$ decays, and $\psi^\prime$ decays in run~I and from
prompt production in run~II.

We now describe how to evaluate the differential hadronic cross section from
Eq.~(\ref{eq:KT}) in combination with the squared matrix elements of the
$2\to1$ and $2\to2$ subprocesses~(\ref{eq:RRtoH}) and (\ref{eq:RRtoHG}),
respectively.
The rapidity and pseudorapidity of a charmonium state with four-momentum
$p^\mu=(p^0,{\bf p}_{T},p^3)$ are given by
\begin{equation}
y=\frac{1}{2}\ln\frac{p^0+p^3}{p^0-p^3},\quad
\eta=\frac{1}{2}\ln\frac{|{\bf p}|+p^3}{|{\bf p}|-p^3},
\end{equation}
respectively.
For the $2\to1$ subprocess~(\ref{eq:RRtoH}), we have
\begin{eqnarray}
&&\frac{d\sigma^{\mathrm{KT}}(p + \overline{p} \to {\cal H} + X)}
{d|{\bf p}_T|d y}
= \frac{|{\bf p}_T|}{(|{\bf p}_T|^2+M^2)^2} \int{d|{\bf
k}_{1T}|^2}\int{d \varphi_1}
\nonumber\\
&&{}\times
\Phi_p(\xi_1,|{\bf k}_{1T}|^2,\mu^2)
\Phi_{\overline{p}}(\xi_2,|{\bf k}_{2T}|^2,\mu^2)
\overline{|{\cal A}(R
+ R \to {\cal H})|^2},
\end{eqnarray}
where
\begin{equation}
\xi_1=\frac{p^0+p^3}{\sqrt{S}},\qquad
\xi_2=\frac{p^0-p^3}{\sqrt{S}},
\label{eq:xi12}
\end{equation}
and ${\bf k}_{2T}={\bf p}_{T}-{\bf k}_{1T}$.
In our numerical analysis, we choose the factorization scale to be
$\mu=M_T$.
For the $2 \to 2$ subprocess~(\ref{eq:RRtoHG}), we have
\begin{eqnarray}
&&\frac{d\sigma^{\mathrm{KT}}(p + \overline{p} \to {\cal H} + X)}{d|{\bf
p}_T| d y}=\frac{|{\bf p}_T|}{(2 \pi)^3} \int{d|{\bf
k}_{1T}|^2}\int{d \varphi_1} \int{d x_2}\int{d|{\bf
k}_{2T}|^2}\int{d \varphi_2}
\nonumber\\
&&{} \times \Phi_p(x_1,|{\bf
k}_{1T}|^2,\mu^2) \Phi_{\overline{p}}(x_2,|{\bf k}_{2T}|^2,\mu^2)
\frac{\overline{|{\cal A}(R + R \to {\cal H} + g)|^2}}{
(x_2 - \xi_2)(2 x_1 x_2 S)^2},
\end{eqnarray}
where
\begin{equation}
x_1=\frac{1}{(x_2 - \xi_2) S}\left[({\bf k}_{1T}+{\bf k}_{2T} - {\bf p}_T)^2-
M^2 - |{\bf p}_{T}|^2 + x_2 \xi_1 S\right].
\end{equation}

We now present and discuss our results. In Table~\ref{tab:NME}, we
list out fit results for the relevant color-octet NMEs for three
different choices of unintegrated gluon distribution function,
namely JB \cite{JB}, JS \cite{JS}, and KMR \cite{KMR}. The
color-singlet NMEs are not fitted, but determined from the measured
partial decay widths of $\psi(nS) \to l^+ + l^-$ and $\chi_{c2} \to
\gamma + \gamma$. The numerical values are adopted from
Ref.~\cite{BKLee} and read: $\langle{\cal
O}^{J/\psi}[^3S_1^{(1)}]\rangle = 1.3$~GeV$^3$, $\langle{\cal
O}^{\psi^\prime}[^3S_1^{(1)}]\rangle = 6.5\times10^{-1}$~GeV$^3$,
and $\langle{\cal O}^{\chi_{cJ}}[^3P_J^{(1)}]\rangle = (2 J +
1)\times 8.9\times 10^{-2}$~GeV$^5$. They were obtained using the
vacuum saturation approximation and heavy-quark spin symmetry in the
NRQCD factorization formulas and including NLO QCD radiative
corrections \cite{QCDCorrections}. The relevant branching ratios are
taken from Ref.~\cite{PDG2004} and read $B(J/\psi \to \mu^+ +
\mu^-)=0.0601$, $B(\psi^\prime \to J/\psi + X)=0.576$, $B(\chi_{c0}
\to J/\psi + \gamma)=0.012$, $B(\chi_{c1} \to J/\psi +
\gamma)=0.318$, and $B(\chi_{c2} \to J/\psi + \gamma)=0.203$. They
somewhat differ from the values used previously \cite{PDG2002}. For
comparison, we list in Table~\ref{tab:NME} also the NMEs obtained in
Ref.~\cite{BKLee} for the collinear parton model with the LO parton
distribution functions of the proton by Martin, Roberts, Stirling,
and Thorne (MRST98LO) \cite{MRST}.

\begin{figure}[ht]
\begin{center}
\includegraphics[width=0.9\textwidth, clip=]{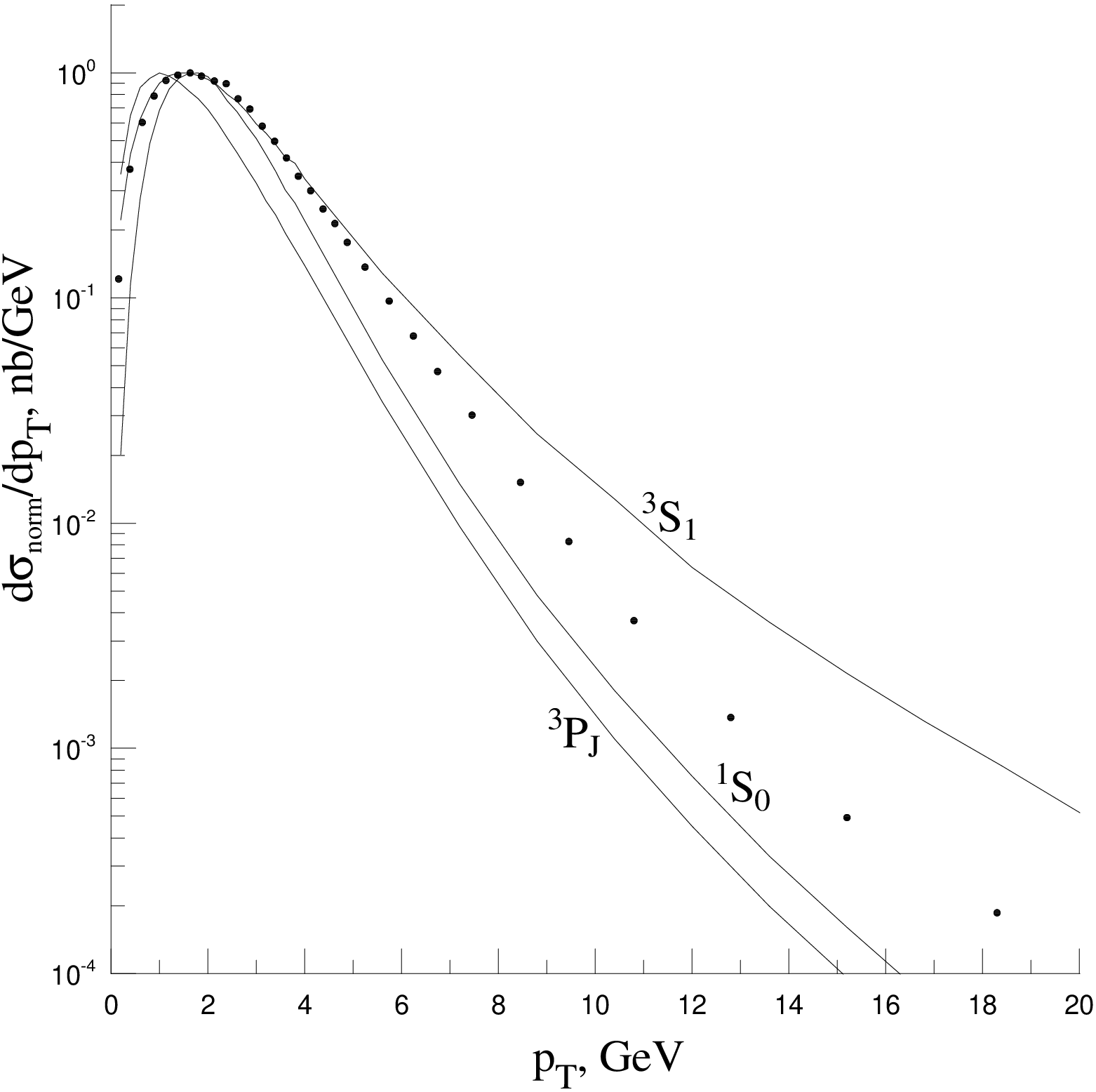}
\end{center}
\caption{\label{fig:States}Contributions to the $p_T$ distribution of prompt
$J/\psi$ hadroproduction in $p\overline{p}$ scattering with
$\sqrt{S}=1.96$~TeV and $|y|<0.6$ from the relevant color-octet states
compared with CDF data from Tevatron run~II \cite{CDFII}.
All distributions are normalized to unity at their peaks.}
\end{figure}

\begin{figure}[ht]
\begin{center}
\includegraphics[width=0.8\textwidth, clip=]{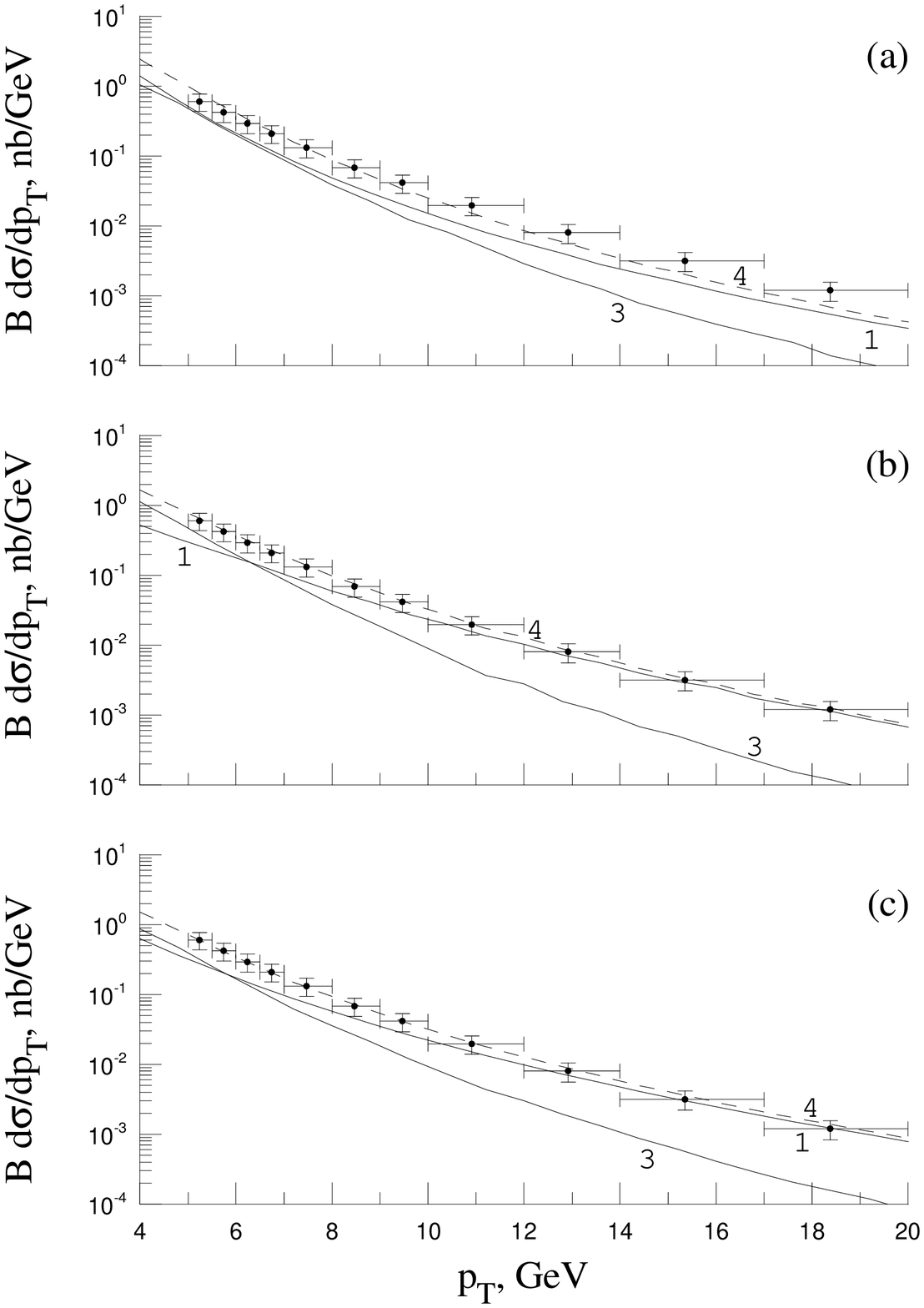}
\end{center}
\caption{Contributions to the $p_T$ distribution of $J/\psi$ mesons from
$\psi^\prime$ decays in hadroproduction in $p\overline{p}$ scattering with
$\sqrt{S}=1.8$~TeV and $|y|<0.6$ from the partonic subprocesses
(1) $R + R\rightarrow \psi^\prime[{^3S}_1^{(8)}]$,
(2) $R + R\rightarrow \psi^\prime[{^1S}_0^{(8)},{^3P}_J^{(8)}]$ (this
contribution actually vanished),
(3) $R + R\rightarrow \psi^\prime[{^3S}_1^{(1)}]+g$,
and (4) their sum
compared with CDF data from Tevatron run~I \cite{CDFI}.
The theoretical results are obtained with the (a) JB \cite{JB}, (b) JS
\cite{JS}, or (c) KMR \cite{KMR} unintegrated gluon distribution functions.
The decay branching fraction $B(J/\psi \to \mu^+ + \mu^-)$ is included.}
\label{fig:Psi2S}
\end{figure}

We first study the relative importance of the different intermediate states in
direct $J/\psi$ and $\psi^\prime$ production.
In previous fits to CDF data from run~I \cite{CDFI}, with $p_T>5$~GeV,
the linear combinations
\begin{equation}
M^{\cal H}_r=
\langle {\cal O}^{\cal H}[^1S_0^{(8)}]\rangle
+\frac{r}{m_c^2}\langle {\cal O}^{\cal H}[^3P_0^{(8)}]\rangle
\label{eq:lc}
\end{equation}
for ${\cal H}=J/\psi,\psi^\prime$ were fixed because it was
infeasible to separate the contributions proportional to $\langle
{\cal O}^{\cal H}[^1S_0^{(8)}]\rangle$ and $\langle {\cal O}^{\cal
H}[^3P_0^{(8)}]\rangle$. By contrast, the new run-II data
\cite{CDFI}, which reach down to $p_T=0$, allow us to determine
$\langle {\cal O}^{\cal H}[^1S_0^{(8)}]\rangle$ and $\langle {\cal
O}^{\cal H}[^3P_0^{(8)}]\rangle$ separately because the respective
contributions exhibit different $p_T$ dependences for $p_T<5$~GeV.
This feature is nicely illustrated in Fig.~\ref{fig:States}, where
the shapes of the relevant color-octet contributions to  prompt
$J/\psi$ production, proportional to $\langle {\cal O}^{\cal
H}[^3S_1^{(8)}]\rangle$, $\langle {\cal O}^{\cal
H}[^1S_0^{(8)}]\rangle$, and $\langle {\cal O}^{\cal
H}[^3P_0^{(8)}]\rangle$, are compared with that of the CDF data
from run~II \cite{CDFII}. Notice that the color-octet
contributions differ in the peak position, by up to 1~GeV.
Apparently, this suffices to disentangle the contributions
previously combined by Eq.~(\ref{eq:lc}). We find that $\langle
{\cal O}^{J/\psi,\psi^\prime}[^3P_0^{(8)}]\rangle$ and $\langle
{\cal O}^{\psi^\prime}[^1S_0^{(8)}]\rangle$ are compatible with
zero, independent of the choice of unintegrated gluon density---a
striking result. For the case of $J/\psi$ production from
$\psi^\prime$ decay, this implies that the $^3S_1^{(1)}$ and
$^3S_1^{(8)}$ channels are sufficient to describe the measured
$p_T$ distribution (see Fig.~\ref{fig:Psi2S}).

\begin{figure}[ht]
\begin{center}
\includegraphics[width=0.8\textwidth, clip=]{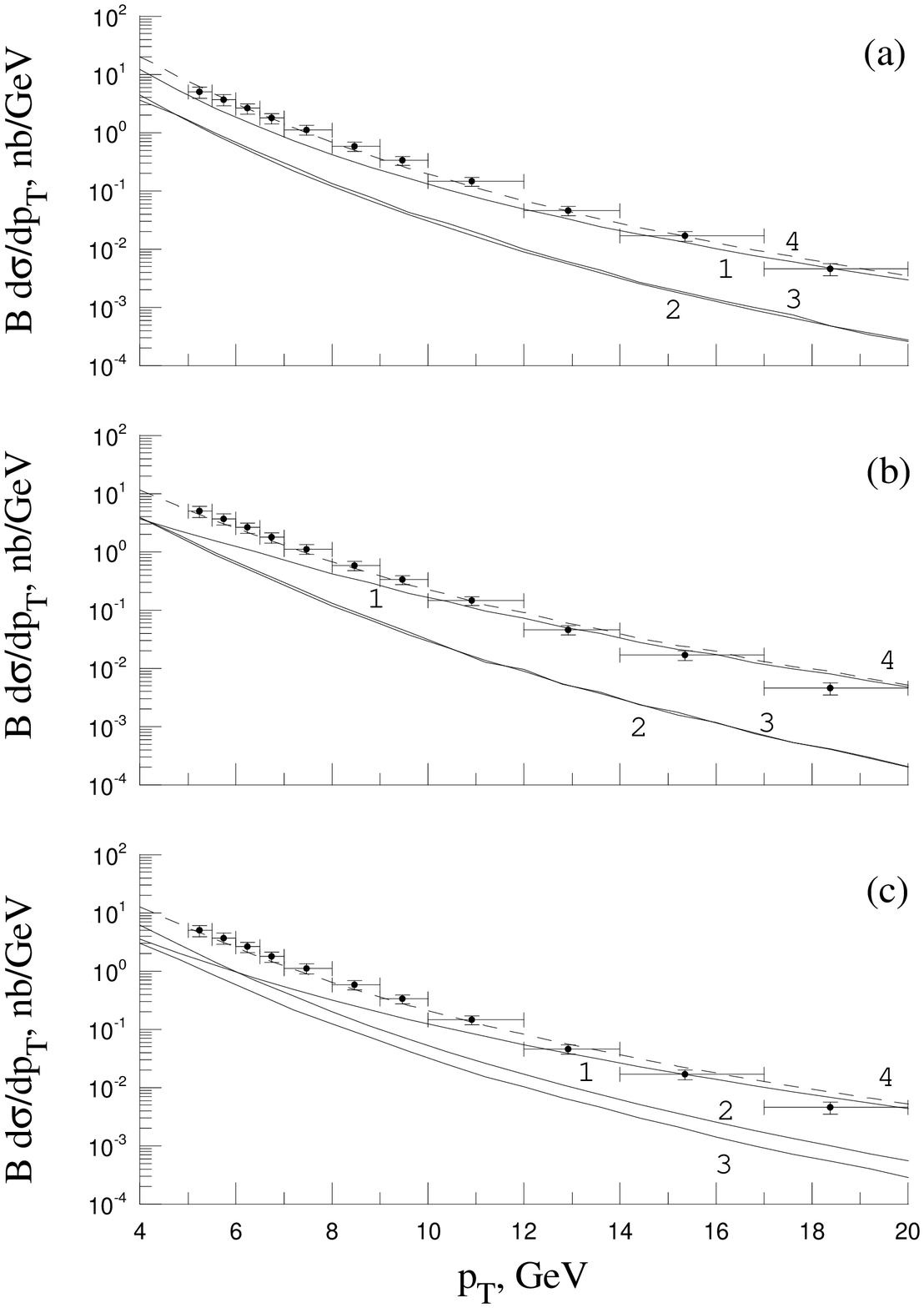}
\end{center}
\caption{\label{fig:Direct}Contributions to the $p_T$ distribution of direct
$J/\psi$ hadroproduction in $p\overline{p}$ scattering with $\sqrt{S}=1.8$~TeV
and $|y|<0.6$ from the partonic subprocesses
(1) $R + R\rightarrow J/\psi[^{3}S_1^{(8)}]$,
(2) $R + R\rightarrow J/\psi[^{1}S_0^{(8)},^{3}P_J^{(8)}]$,
(3) $R + R\rightarrow J/\psi[^{3}S_1^{(1)}]+g$,
and (4) their sum
compared with CDF data from Tevatron run~I \cite{CDFI}.
The theoretical results are obtained with the (a) JB \cite{JB}, (b) JS
\cite{JS}, or (c) KMR \cite{KMR} unintegrated gluon distribution functions.
The decay branching fraction $B(J/\psi \to \mu^+ + \mu^-)$ is included.}
\end{figure}

\begin{figure}[ht]
\begin{center}
\includegraphics[width=1.0\textwidth, clip=]{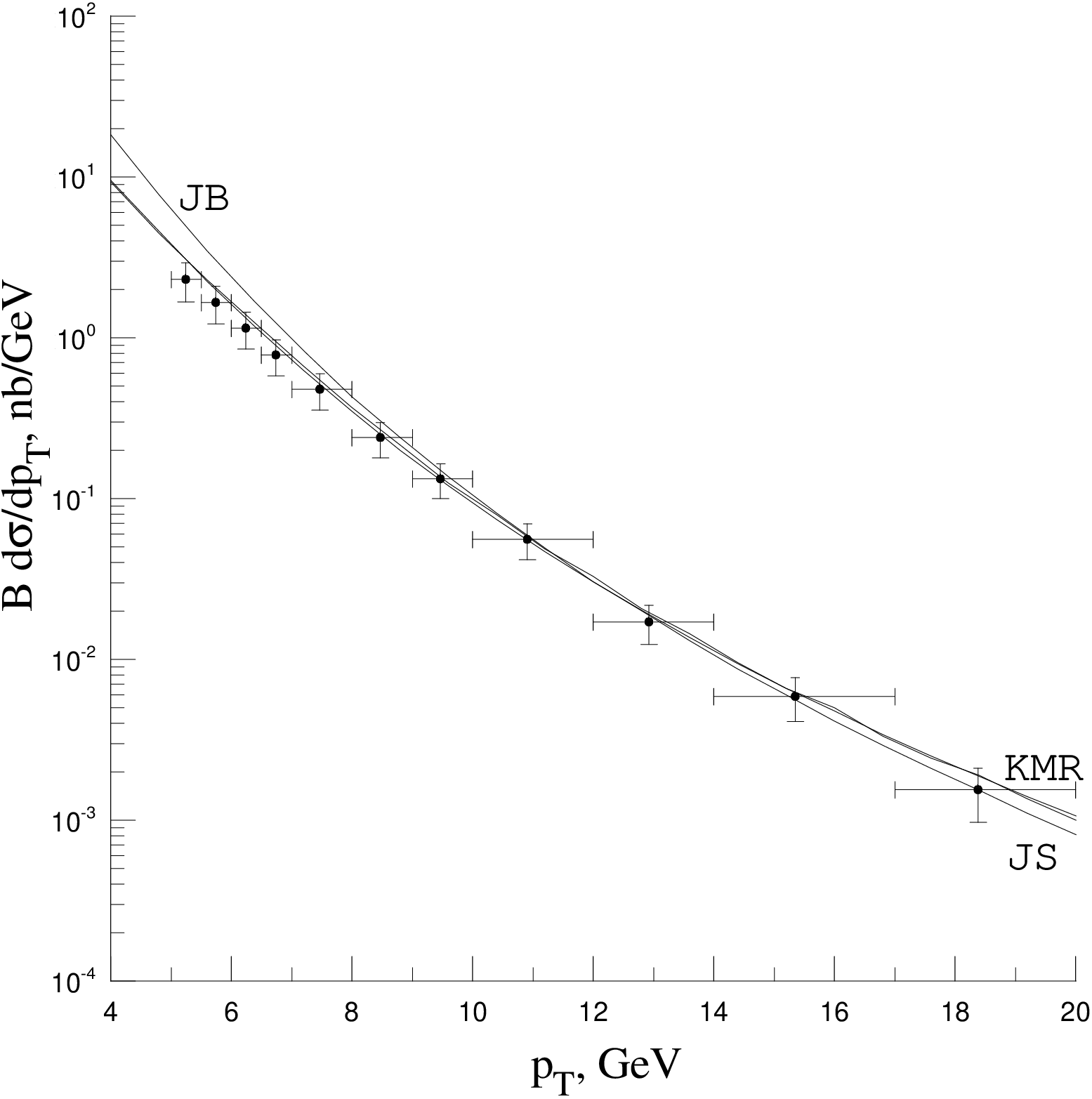}
\end{center}
\caption{\label{fig:ChiCJ}Contributions to the $p_T$ distribution of $J/\psi$
mesons from $\chi_{cJ}$ decays in hadroproduction in $p\overline{p}$
scattering with $\sqrt{S}=1.8$~TeV and $|y|<0.6$ from the sum of the partonic
subprocesses $R + R\rightarrow \chi_{cJ}[^{3}P_J^{(1)}]$ and
$R + R\rightarrow \chi_{cJ}[^{3}S_1^{(8)}]$, the latter of which being quite
unimportant,
compared with CDF data from Tevatron run~I \cite{CDFI}.
The theoretical results are obtained with the JB \cite{JB}, JS
\cite{JS}, or KMR \cite{KMR} unintegrated gluon distribution functions.
The decay branching fraction $B(J/\psi \to \mu^+ + \mu^-)$ is included.}
\end{figure}

\begin{figure}[ht]
\begin{center}
\includegraphics[width=0.8\textwidth, clip=]{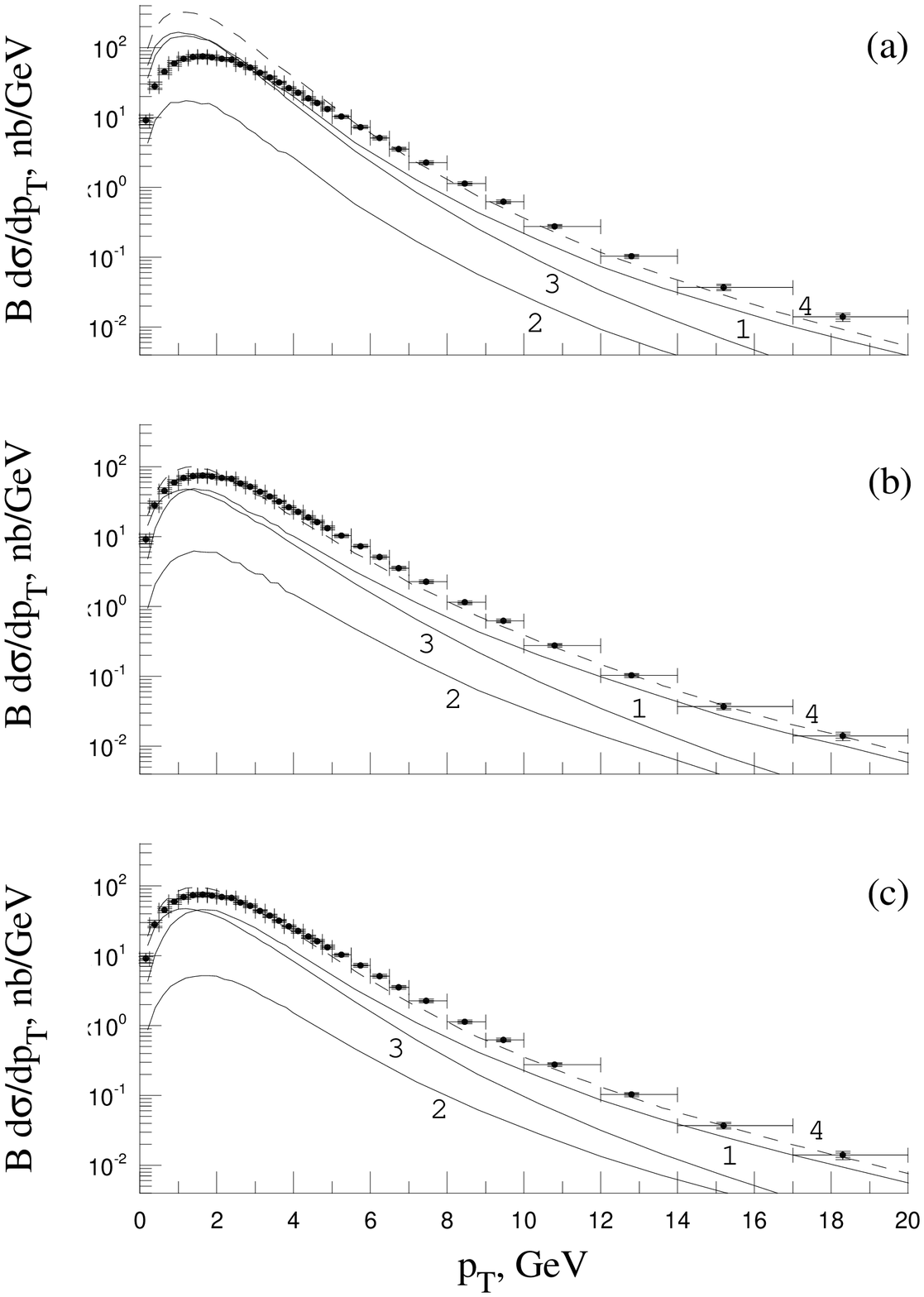}
\end{center}
\caption{\label{fig:PromptR2}Contributions to the $p_T$ distribution of prompt
$J/\psi$ hadroproduction in $p\overline{p}$ scattering with
$\sqrt{S}=1.96$~TeV and $|y|<0.6$ from
(1) direct production,
(2) $\psi^\prime$ decays,
(3) $\chi_{cJ}$ decays,
and (4) their sum
compared with CDF data from Tevatron run~II \cite{CDFII}.
The theoretical results are obtained with the (a) JB \cite{JB}, (b) JS
\cite{JS}, or (c) KMR \cite{KMR} unintegrated gluon distribution functions.
The decay branching fraction $B(J/\psi \to \mu^+ + \mu^-)$ is included.}
\end{figure}

In Figs.~\ref{fig:Direct}--\ref{fig:PromptR2}, we compare the CDF data on
$J/\psi$ mesons from direct production, $\psi^\prime$ decays, and $\chi_{cJ}$
decays in run I \cite{CDFI} and from prompt production in run II \cite{CDFII},
respectively, with the theoretical results evaluated with the NMEs listed in
Table~\ref{tab:NME}.
From Fig.~\ref{fig:Direct}, we observe that the color-singlet contribution is
significant, especially at low values of $p_T$, and comparable to the one
from the $^1S_0^{(8)}$ channel.
As is familiar from the collinear parton model, the $^3S_1^{(8)}$ contribution
makes up the bulk of the cross section at large values of $p_T$.
Incidentally, the values of $\langle {\cal O}^{J/\psi}[^3S_1^{(8)}]\rangle$
obtained in the $k_T$-factorization framework are in average quite close to
the one obtained in the collinear parton model, as may be seen from
Table~\ref{tab:NME}.
The situation is very similar for $J/\psi$ production from $\psi^\prime$
decay, considered in Fig.~\ref{fig:Psi2S}, except that the
$^1S_0^{(8)}$ and $^3P_J^{(8)}$ contributions are negligible.

At this point, we wish to compare our results for direct $J/\psi$
hadroproduction in the $k_T$-factorization approach with the literature,
specifically with Refs.~\cite{CDFBaranov,KTYuan}, which consider the partonic
subprocess~(\ref{eq:RRtoH}).
By contrast, in Ref.~\cite{KTTeryaev}, the NLO subprocess
$R + R \to J/\psi[{^3}S_1^{(8)}] + g$ was studied, leaving aside the LO
subprocess~(\ref{eq:RRtoH}).
In Ref.~\cite{KTYuan}, the value
$\langle{\cal O}^{J/\psi}[{}^3S_1^{(8)}]\rangle = 7.0\times 10^{-3}$~GeV$^3$
was obtained using the Kwiecinski-Martin-Stasto (KMS) \cite{KMS} unintegrated
gluon distribution function.
This value is 2.6 times larger than the result we found using the KMR
\cite{KMR} version, which is very similar to the KMS one.
We attribute this difference in
$\langle{\cal O}^{J/\psi}[{}^3S_1^{(8)}]\rangle$ to the different scale choice,
$\mu=k_T$, used by the authors of Ref.~\cite{KTYuan}.
Adopting their value for $\langle{\cal O}^{J/\psi}[{}^3S_1^{(8)}]\rangle$, we
can reproduce their result for the respective cross section contribution.
On the other hand, the value
$\langle{\cal O}^{J/\psi}[{}^3S_1^{(8)}]\rangle=
15.0\times 10^{-3}$~GeV$^3$ found in Ref.~\cite{CDFBaranov} exceeds the one of
Ref.~\cite{KTYuan} by a factor of 2.1 and our KMR value by a factor of 5.6.
Furthermore, the cross section evaluated in Ref.~\cite{CDFBaranov} falls off
with $p_T$ considerably more slowly than in Ref.~\cite{KTYuan} and here, only
by one order of magnitude as $p_T$ runs from 2 to 20~GeV, while the
unintegrated gluon density in the proton falls off with $k_T$ far more rapidly.

The discussion of $J/\psi$ production from radiative $\chi_{cJ}$ decays,
considered in Fig.~\ref{fig:ChiCJ}, is simpler because there is only one free
parameter in the fit, namely
$\langle{\cal O}^{\chi_{c0}}[{^3}S_1^{(8)}]\rangle$.
We confirm the conclusion of Ref.~\cite{KTTeryaev}, that, in the
$k_T$-factorization approach, the color-singlet contribution is sufficient to
describe the data.
In fact, the best fit is realized when
$\langle{\cal O}^{\chi_{c0}}[{^3}S_1^{(8)}]\rangle$ is taken to be zero or
very small.
In case of the JB gluon density, the fitting procedure even favors a
negative value of $\langle{\cal O}^{\chi_{c0}}[{^3}S_1^{(8)}]\rangle$.

In Fig.~\ref{fig:PromptR2}, the $p_T$ distribution of prompt $J/\psi$
production in run~II is broken down into the contributions from direct
production, $\psi^\prime$ decays, and $\chi_{cJ}$ decays.
We observe that the latter is dominant for $p_T \alt 5$~GeV, while prompt
$J/\psi$ mesons are preferably produced directly at larger values of $p_T$.
The contribution from $\psi^\prime$ decays stays at the level of several
percent for all values of $p_T$.
While the JS \cite{JS} and KMR \cite{KMR} gluon densities allow for a
faithful description of the measured $p_T$ distribution \cite{CDFII}, the JB
\cite{JB} one has a problem in the low-$p_T$ range, at $p_T \alt 5$~GeV, where
even the $\chi_{cJ}$-decay contribution, which is entirely of color-singlet
origin, exceeds the data.
This problem can be traced to the speed of growth of the JB gluon density as
$k_T \to 0$.
By contrast, the JS and KMR gluon densities are smaller and approximately
$k_T$ independent at low values of $k_T$.
For this reason, we excluded the CDF prompt-$J/\psi$ data from run~I
\cite{CDFI} and run~II \cite{CDFII} from our fit based on the JB gluon
density.

Considering the color-octet NMEs relevant for the $J/\psi$,
$\psi^\prime$ and $\chi_{cJ}$ production mechanisms, we can
formulate the following heuristic rule for favoured transitions
from color-octet to color-singlet states: $\Delta L\simeq 0$ and
$\Delta S\simeq 0$; {\it i.e.} these transitions are doubly
chromoelectric and preserve the orbital angular momentum and the
spin of the heavy-quark bound state.

\section{\label{sec:six}Charmonium production at HERA}

At HERA, the cross section of prompt $J/\psi$ production was
measured in a wide range of the kinematic variables
$W^2=(P+q)^2$, $Q^2=-q^2$, $y_2=(P\cdot q)/(P\cdot k)$,
$z=(P\cdot p)/(P\cdot q)$, $p_T$ and $y$, where $P^\mu$, $k^\mu$,
$k^{\prime\mu}$, $q^\mu=k^\mu-k^{\prime\mu}$, and $p^\mu$ are the
four-momenta of the incoming proton, incoming lepton, scattered
lepton, virtual photon, and produced $J/\psi$ meson,
respectively, both in photoproduction \cite{epZEUS}, at small
values of $Q^2$, and deep-inelastic scattering (DIS) \cite{epH1},
at large values of $Q^2$. At sufficiently large values of $Q^2$,
the virtual photon behaves like a point-like object, while, at
low values of $Q^2$, it can either act as a point-like object
(direct photoproduction) or interact via its quark and gluon
content (resolved photoproduction). Resolved photoproduction is
only important at low values of $z$.

In the region $z\alt 1$, diffractive production, which is beyond the
scope of this paper, takes place.
In order to suppress the diffractive-production contribution, one usually
applies the acceptance cut $z<0.9$.
This effectively eliminates the contributions from the $2\to1$ partonic
subprocesses~(\ref{eq:FRtoH}) and (\ref{eq:eRtoH}), so that we are left with
the $2\to2$ partonic subprocesses~(\ref{eq:FRtoHG}) and (\ref{eq:eRtoHG}).

Let us first present the relevant formulas for the double differential
cross sections of DIS, direct photoproduction, and resolved photoproduction.
In the case of DIS, we have
\begin{eqnarray}
&&\frac{d\sigma^{\mathrm{KT}}(p + e \to e+{\cal H} + X)}
{d|{\bf p}_T|^2 d z}
=\frac{1}{8 z (2 \pi)^5}\int{d Q^2}\int{d y_2}
\int{d|{\bf k}_{1T}|^2}\int{d\varphi_1}\int{d \varphi_2}
\nonumber\\
&&{}\times \Phi_p(x_1,|{\bf k}_{1T}|^2,\mu^2)
\frac{\overline{|{\cal A}(R + e \to e + {\cal H} + g)|^2}}
{(y_2 - \chi_2) (2 x_1 S)^2},
\end{eqnarray}
where
\begin{eqnarray}
x_1&=&\frac{1}{(y_2 -
\chi_2) S}\left[({\bf k}_{1T}+{\bf q}_{2T}- {\bf p}_{T})^2 - M^2 -
|{\bf p}_{T}|^2 + y_2 \chi_1 S + (y_2 - \chi_2) Q^2\right],
\nonumber\\
\chi_1&=&\frac{p^0+p^3}{2 E_p},\qquad
\chi_2=\frac{p^0-p^3}{2 E_e}.
\end{eqnarray}
Here, $E_p$ and $E_e$ are the proton and lepton energies in the laboratory
frame, and we have $S=4 E_p E_e$ and $|{\bf q}_{2T}|=\sqrt{(1-y_2)Q^2}$.

In the case of direct photoproduction, we have
\begin{eqnarray}
\lefteqn{\frac{d\sigma^{\mathrm{KT}}(p + e \to e+{\cal H} + X)}{d|{\bf
p}_T|^2 d z}=\frac{1}{2 z (2 \pi)^2}\int{d y_2}
 \int{d|{\bf k}_{1T}|^2}\int{d \varphi_1}}
\nonumber\\
&&{}\times\Phi_p(x_1,|{\bf k}_{1T}|^2,\mu^2) f_{\gamma/e}(y_2)
\frac{\overline{|{\cal A}(R + \gamma \to {\cal H} + g)|^2}}
{y_2(y_2 - \chi_2)(2 x_1S)^2},
\end{eqnarray}
where
\begin{equation}
x_1=\frac{1}{(y_2 - \chi_2) S}\left[({\bf k}_{1T}-{\bf p}_{T})^2
- M^2 - |{\bf p}_{T}|^2 + y_2 \chi_1 S\right]
\end{equation}
and $f_{\gamma/e}(y_2)$ is the quasi-real photon flux.
In the Weiz\"acker-Williams approximation, the latter takes the form
\begin{equation}
f_{\gamma/e}(y_2)=\frac{\alpha}{2\pi}\left[\frac{1+(1-y_2)^2}{y_2}\ln
\frac{Q_{\rm max}^2}{Q_{\rm min}^2}
+2m_e^2y_2\left(\frac{1}{Q_{\rm min}^2}-\frac{1}{Q_{\rm max}^2}\right)\right],
\end{equation}
where $Q_{\rm min}^2=m_e^2y_2^2/(1-y_2)$ and $Q^2_{\rm max}$ is determined by
the experimental set-up, {\it e.g.}\
$Q^2_{\rm max}=1$~GeV$^2$ \cite{epZEUS}.

In the case of resolved photoproduction, we take into account the $2 \to 1$
and $2\to 2$ partonic subprocesses (\ref{eq:RRtoH}) and (\ref{eq:RRtoHG}),
respectively, where the first reggeized gluon comes from the proton and the
second one from the photon.
For subprocess (\ref{eq:RRtoH}), the relevant doubly differential cross
section reads:
\begin{eqnarray}
\lefteqn{\frac{d\sigma^{\mathrm{KT}}(p + e \to e+{\cal H} + X)}{d|{\bf
p}_T|^2 d z}=\frac{1}{2 z (|{\bf p}_T|^2 +M^2)^2} \int{d y_2}
\int{d|{\bf k}_{1T}|^2}\int{d \varphi_1}}
\nonumber\\
&&\times \Phi_p(x_1,|{\bf k}_{1T}|^2,\mu^2)f_{\gamma/e}(y_2)
\Phi_\gamma(x_2,|{\bf k}_{2T}|^2,\mu^2)
{\overline{|{\cal A}(R + R \to {\cal H})|^2}},
\end{eqnarray}
where
\begin{eqnarray}
x_1=\chi_1, \quad x_2=\frac{\chi_2}{y_2}, \quad {\bf k}_{2T}={\bf
p}_{T}-{\bf k}_{1T}.
\end{eqnarray}
For subprocess (\ref{eq:RRtoHG}), the relevant doubly differential cross
section is given by
\begin{eqnarray}
&&\frac{d\sigma^{\mathrm{KT}}(p + e \to e+{\cal H} + X)}{d|{\bf
p}_T|^2 d z}=\frac{1}{2 z (1 - z) (2 \pi)^3} \int{d y_2}\int{d|{\bf
k}_{1T}|^2}\int{d \varphi_1}\int{d x_2}\int{d|{\bf k}_{2T}|^2}\int{d
\varphi_2}
\nonumber\\
&&\times \Phi_p(x_1,|{\bf k}_{1T}|^2,\mu^2)f_{\gamma/e}(y_2)
\Phi_\gamma(x_2,|{\bf k}_{2T}|^2,\mu^2)
\frac{\overline{|{\cal A}(R + R \to {\cal H} +
g)|^2}}{x_2 (2 x_1 x_2 y_2 S)^2},
\end{eqnarray}
where
\begin{eqnarray}
x_1&=&\frac{1}{(x_2 y_2 - \chi_2) S}\left[({\bf k}_{1T}-{\bf
p}_{T})^2 - M^2 - |{\bf p}_{T}|^2 + x_2 y_2 \chi_1 S\right].
\end{eqnarray}
To evaluate the unintegrated gluon distribution function in the resolved
photon, $\Phi_\gamma(x_2,|{\bf k}_{2T}|^2,\mu^2)$, we use a procedure
suggested by Bl\"umlein \cite{JBPhoton}, which is similar to the proton case
\cite{JB}.
As input for this, we use the collinear parton distribution functions of the
resolved photon by Gl\"uck, Reya, and Vogt (GRV$_\gamma$) \cite{GRVPhoton}.

\begin{figure}[ht]
\begin{center}
\includegraphics[width=1.0\textwidth, clip=]{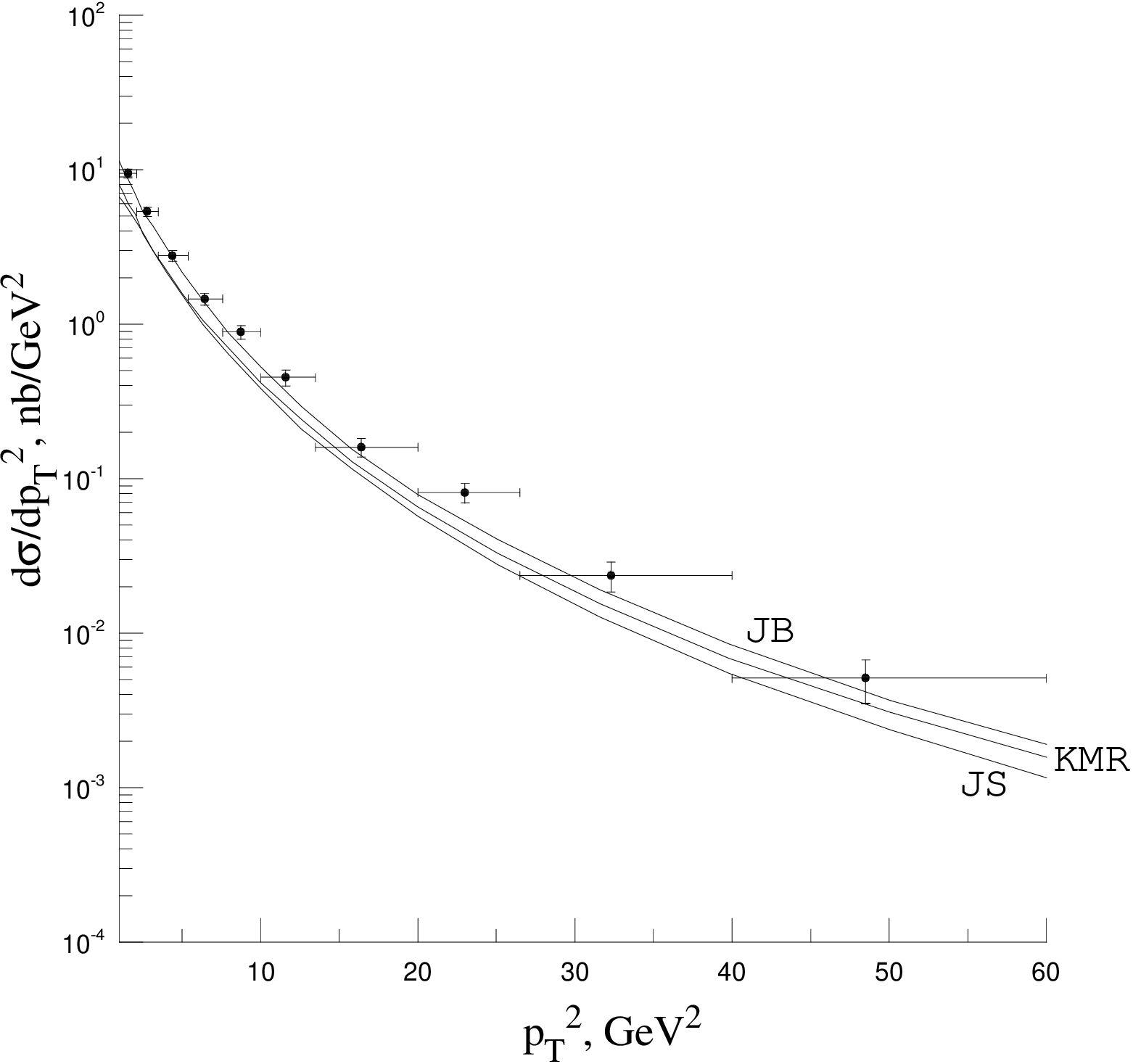}
\end{center}
\caption{\label{fig:PhP}Contribution to the $p_T^2$ distribution of prompt
$J/\psi$ photoproduction in $ep$ scattering with $E_p=820$~GeV,
$E_e=27.5$~GeV, $60$~GeV${}<W<240$~GeV, $Q^2<1$~GeV$^2$, and $0.3<z<0.9$ from
the direct-photon subprocess
$R + \gamma\rightarrow {\cal H}[^{3}S_1^{(1)}]+g$
compared with ZEUS data from HERA \cite{epZEUS}.
The resolved-photon subprocesses
$R + R\rightarrow {\cal H}[^{3}S_1^{(1)},^{3}P_J^{(1)},^{3}S_1^{(8)},
^{1}S_0^{(8)},^{3}P_J^{(8)}]$
are neglected.
The theoretical results are obtained with the JB \cite{JB}, JS \cite{JS}, or
KMR \cite{KMR} unintegrated gluon distribution functions.}
\end{figure}

\begin{figure}[ht]
\begin{center}
\includegraphics[width=0.8\textwidth, clip=]{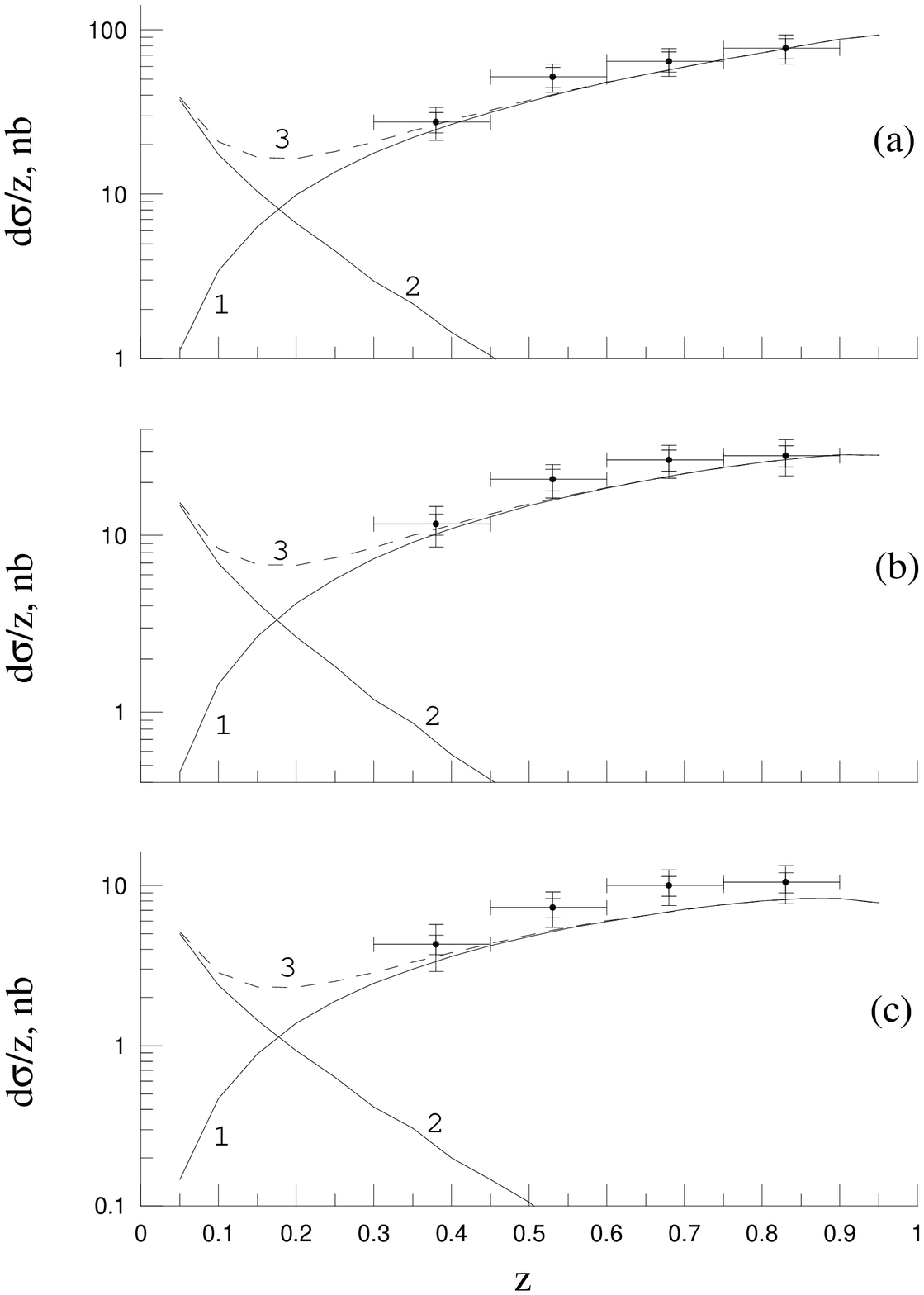}
\end{center}
\caption{\label{fig:PhPz}Contributions to the $z$ distribution of prompt
$J/\psi$ photoproduction in $ep$ scattering with $E_p=820$~GeV,
$E_e=27.5$~GeV, 60~GeV${}<W<240$~GeV, $Q^2<1$~GeV$^2$, and (a) $p_T>1$~GeV,
(b) $p_T>2$~GeV, or (c) $p_T>3$~GeV from
(1) the direct-photon subprocess
$R + \gamma\rightarrow {\cal H}[^{3}S_1^{(1)}]+g$,
(2) the resolved-photon subprocesses
$R + R\rightarrow {\cal H}[^{3}S_1^{(1)},^{3}P_J^{(1)},^{3}S_1^{(8)},
^{1}S_0^{(8)},^{3}P_J^{(8)}]$, and
(3) their sum
compared with ZEUS data from HERA \cite{epZEUS}.
The theoretical results are obtained with the JB \cite{JB}
unintegrated gluon distribution function.}
\end{figure}

\begin{figure}[ht]
\begin{center}
\includegraphics[width=1.0\textwidth, clip=]{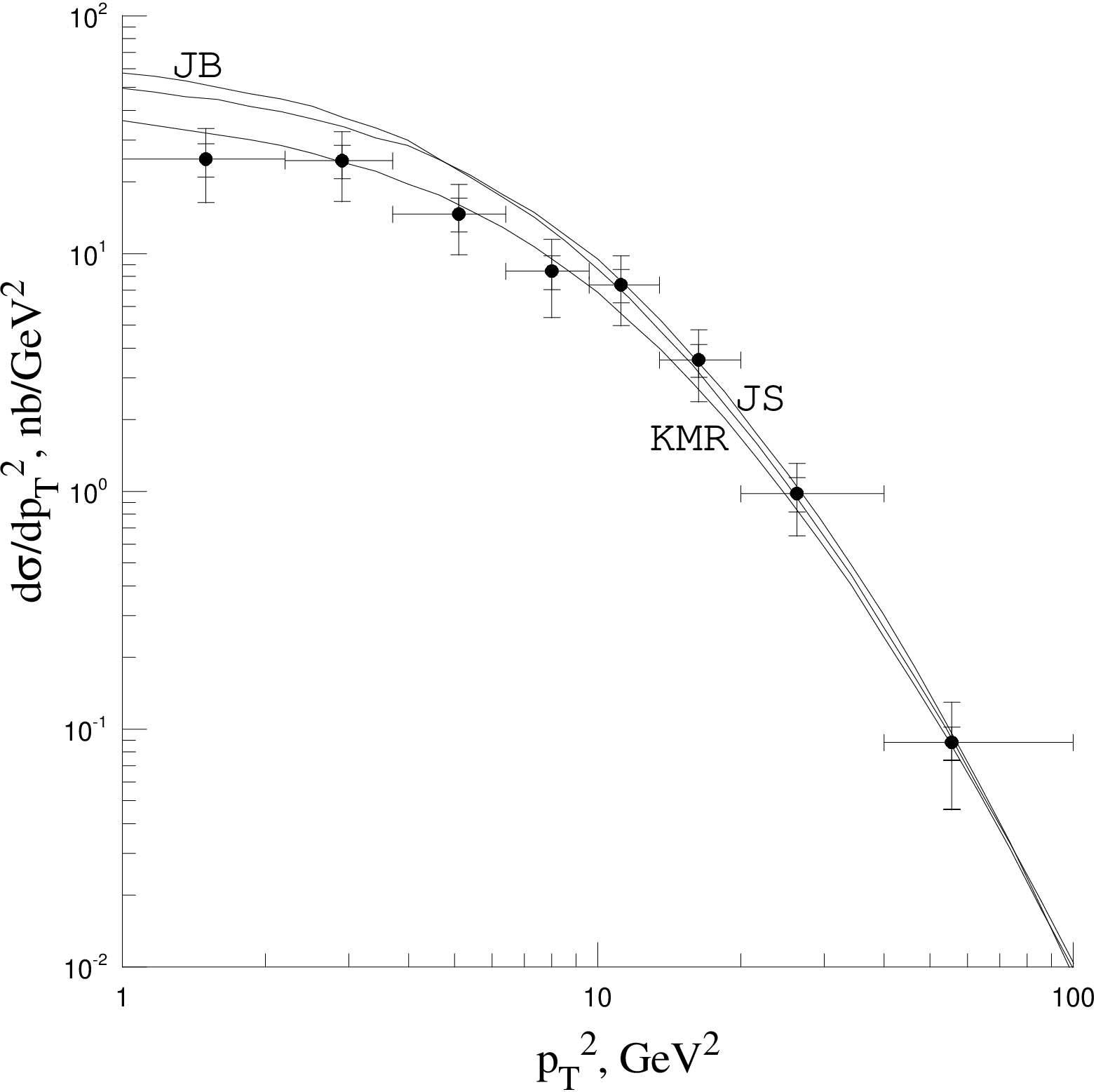}
\end{center}
\caption{\label{fig:DIS}Contribution to the $p_T^2$ distribution of prompt
$J/\psi$ electroproduction in $ep$ scattering with $E_p=920$~GeV,
$E_e=27.5$~GeV, 50~GeV${}<W<225$~GeV, 2~GeV$^2<Q^2<100$~GeV$^2$, and
$0.3<z<0.9$ from the color-singlet subprocess
$R + e\rightarrow e + {\cal H}[^{3}S_1^{(1)}]+g$ compared with H1 data from
HERA \cite{epH1}.
The theoretical results are obtained with the JB \cite{JB}, JS \cite{JS}, or
KMR \cite{KMR} unintegrated gluon distribution functions.}
\end{figure}

\begin{figure}[ht]
\begin{center}
\includegraphics[width=1.0\textwidth, clip=]{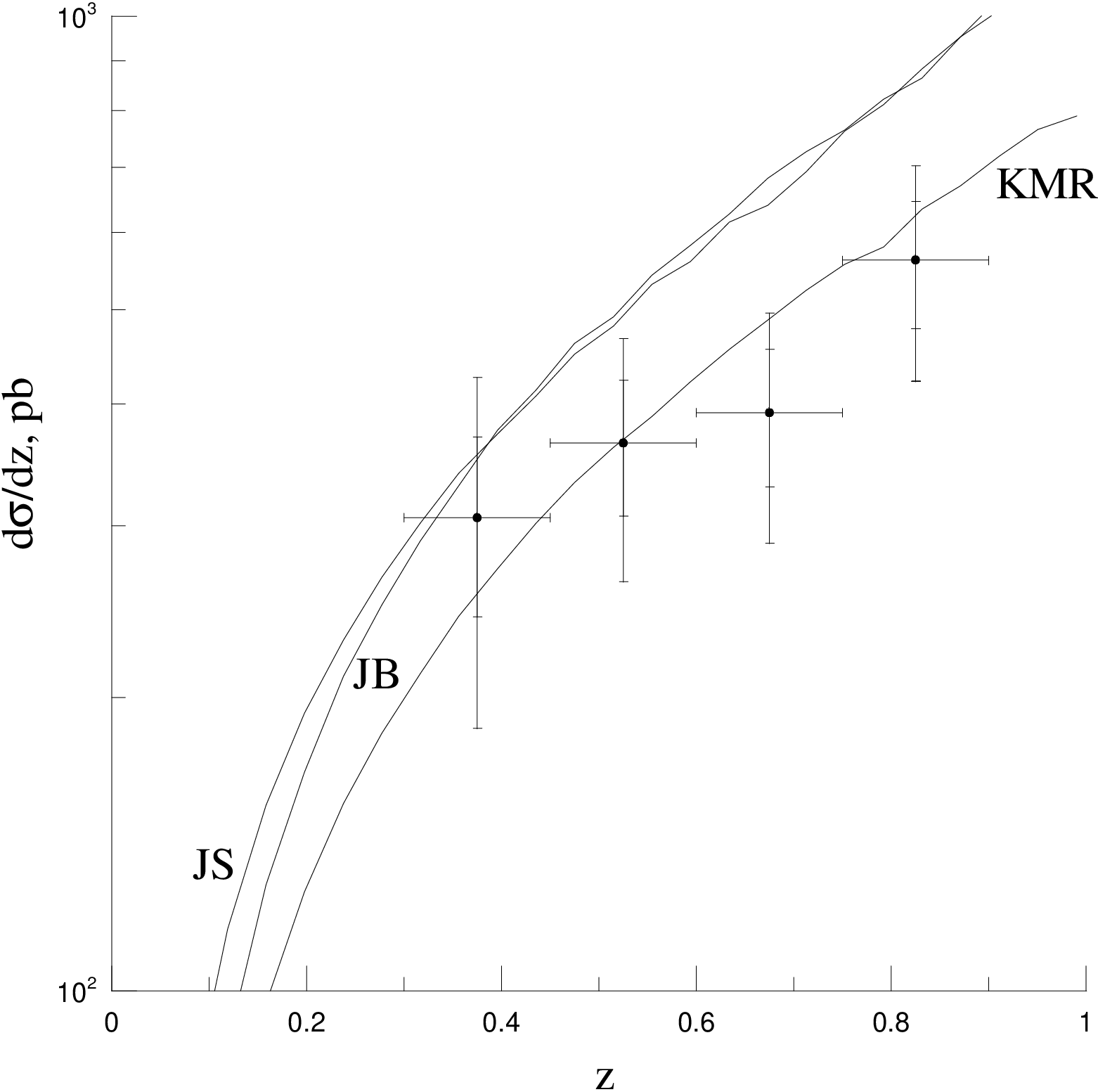}
\end{center}
\caption{\label{fig:DISz}Contribution to the $z$ distribution of prompt
$J/\psi$ electroproduction in $ep$ scattering with $E_p=920$~GeV,
$E_e=27.5$~GeV, 50~GeV${}<W<225$~GeV, 2~GeV$^2<Q^2<100$~GeV$^2$, and
$p_T > 1$~GeV from the color-singlet subprocess
$R + e\rightarrow e + {\cal H}[^{3}S_1^{(1)}]+g$
compared with H1 data from HERA \cite{epH1}.
The theoretical results are obtained with the JB \cite{JB}, JS \cite{JS}, or
KMR \cite{KMR} unintegrated gluon distribution functions.}
\end{figure}

In Figs.~\ref{fig:PhP}--\ref{fig:DISz}, our NRQCD predictions in the
$k_T$-factorization approach, evaluated with the NMEs from
Table~\ref{tab:NME}, are compared with the HERA data \cite{epZEUS,epH1}.
Specifically, Figs.~\ref{fig:PhP} and \ref{fig:PhPz} refer to the $p_T^2$ and
$z$ distributions in photoproduction with $E_p=820$~GeV, $E_e=27.5$~GeV,
60~GeV${}<W<240$~GeV, and $Q^2<1$~GeV$^2$ \cite{epZEUS}, while
Figs.~\ref{fig:DIS} and \ref{fig:DISz} refer to those in DIS with
$E_p=920$~GeV, $E_e=27.5$~GeV, 50~GeV${}<W<225$~GeV, and
2~GeV$^2<Q^2<100$~GeV$^2$ \cite{epH1}.
Acceptance cuts common to both photoproduction and DIS include $p_T>1$~GeV and
$0.3<z<0.9$.
In this regime, the LO NRQCD predictions in the $k_T$-factorization approach
are mainly due to the color-singlet channels and are thus fairly independent
of the color-octet NMEs presented in Table~\ref{tab:NME}.
Therefore, our results agree well with previous calculations in the CSM
\cite{SaleevZotov94}, up to minor differences in the choice of the
color-singlet NMEs and the $c$-quark mass.

\section{\label{sec:seven}Charmonium production at LEP2}

Some time ago, the DELPHI Collaboration presented data on the
inclusive cross section of $J/\psi$ photoproduction in
$\gamma\gamma$ collisions ($e^+ + e^- \to e^+ + e^- + J/\psi +
X$) at LEP2, taken as a function of the $J/\psi$ transverse
momentum $p_T$ \cite{LEPJpsi}. The $J/\psi$ mesons were
identified through their decays to $\mu^+\mu^-$ pairs, and events
where the system $X$ contains a prompt photon were suppressed by
requiring that at least four charged tracks were reconstructed.
The average $e^+e^-$ center-of-mass energy was
$\sqrt{S}=197$~GeV, the scattered positrons and electrons were
antitagged, with maximum angle $\theta_{\rm max}=32$~mrad, and
the maximum $\gamma\gamma$ center-of-mass energy was chosen to be
$W=35$~GeV in order to reject the major part of the
non-two-photon events.

Under LEP2 experimental conditions, most $J/\psi$ mesons are produced
promptly, while the cross section for $J/\psi$ mesons from $b$-hadron decays
is estimated to be about 1\% of the total $J/\psi$ cross section
\cite{KniehlLEP} and can be safely neglected.
Because the average value of the photon virtuality $Q^2$ is small, the
Weizs\"acker-Williams approximation can be used to evaluate the $e^+ e^-$
cross section from the $\gamma\gamma$ cross section as
\begin{equation}
d\sigma(e^+ + e^- \to e^+ + e^- + {\cal H} + X)=\int dy_1\int dy_2\,
f_{\gamma/e}(y_1)f_{\gamma/e}(y_2) d\sigma(\gamma + \gamma \to
{\cal H} + X).
\end{equation}

The process $e^+ + e^- \to e^+ + e^- + J/\psi + X$ receives contributions from
direct, single-resolved, and double-resolved photoproduction.
The relevant partonic subprocesses are:
$\gamma + \gamma \to {\cal H}[{^3S}_1^{(8)}] + g$,
$\gamma + R \to {\cal H}[{^1S}_0^{(8)}, {^3P}_{J}^{(8)}]$,
$\gamma + R \to {\cal H}[{^3S}_1^{(1)}] + g$,
$R + R \to {\cal H}[{^3S}_1^{(8)}, {^1S}_0^{(8)}, {^3P}_{J}^{(8)}]$, and
$R + R \to {\cal H}[{^3S}_1^{(1)}] + g$.
The squared amplitude of $\gamma + \gamma \to {\cal H}[^3S_1^{(8)}] + g$ may
be found in Ref.~\cite{KniehlLEP}, the ones for the other partonic
subprocesses were presented in Sec.~\ref{sec:four}.

The cross section of direct photoproduction is evaluated as
\begin{eqnarray}
&&\frac{d\sigma(e^+ + e^-\to e^+ + e^-+{\cal H} + X)}{d|{\bf p}_T|^2 d y}=
\frac{1}{4 \pi} \int{d y_2}\, f_{\gamma/e}(y_1) f_{\gamma/e}(y_2)
\nonumber\\
&&{}\times\frac{y_1 y_2}{y_2 - \xi_2}\,
\frac{\overline{|{\cal A}(\gamma + \gamma \to {\cal H} + g)|^2}}
{(2 y_1 y_2 S)^2},
\end{eqnarray}
where $\xi_1$ and $\xi_2$ are defined in Eq.~(\ref{eq:xi12}) and
\begin{equation}
y_1=\frac{y_2 \xi_1 S - M^2}{(y_2 - \xi_2) S}.
\end{equation}

In the case of single-resolved photoproduction via the $2\to1$ subprocesses,
we have
\begin{eqnarray}
&&\frac{d\sigma^{\mathrm{KT}}(e^+ + e^- \to e^+ + e^- + {\cal H} + X)}{d|{\bf
p}_T|^2 d y}= 4 \pi \int{d y_1}\, f_{\gamma/e}(y_1)
f_{\gamma/e}(y_2)
\nonumber\\
&&\times \Phi_\gamma(x_1,|{\bf k}_{1T}|^2,\mu^2) y_2
\frac{\overline{|{\cal A}(R + \gamma \to {\cal H} )|^2}}{(2 x_1
y_1 y_2 S)^2},
\end{eqnarray}
where $x_1=\xi_1/y_1$, $y_2=\xi_2$, and
${\bf k}_{1T}={\bf p}_{T}$.
In the case of single-resolved photoproduction via the $2\to2$ subprocess, we
have
\begin{eqnarray}
&&\frac{d\sigma^{\mathrm{KT}}(e^+ + e^- \to e^+ + e^-+{\cal H} + X)}{d|{\bf
p}_T|^2 d y}
=\frac{1}{2 (2 \pi)^2}\int{d y_1} \int{d y_2}
\int{d|{\bf k}_{1T}|^2}\int{d \varphi_1}\, f_{\gamma/e}(y_1) f_{\gamma/e}(y_2)
\nonumber\\
&&{}\times \Phi_\gamma(x_1,|{\bf k}_{1T}|^2,\mu^2) \frac{y_2}{y_2 - \xi_2}\,
\frac{\overline{|{\cal A}(R + \gamma \to {\cal
H} + g)|^2}}{(2 x_1 y_1 y_2 S)^2},
\end{eqnarray}
where
\begin{equation}
x_1=\frac{1}{y_1 (y_2 - \xi_2) S}\left[({\bf k}_{1T}-{\bf
p}_{T})^2 - M^2 - |{\bf p}_{T}|^2 + y_2 \xi_1 S\right].
\end{equation}

In the case of double-resolved photoproduction via the $2\to1$ subprocesses,
we have
\begin{eqnarray}
&&\frac{d\sigma^{\mathrm{KT}}(e^+ + e^- \to  e^+ + e^-+{\cal H} + X)}{d|{\bf
p}_T|^2 d y}= 2\int{d y_1} \int{d y_2}
 \int{d|{\bf k}_{1T}|^2}\int{d \varphi_1}\,
f_{\gamma/e}(y_1) f_{\gamma/e}(y_2)
\nonumber\\
&&{}\times \Phi_\gamma(x_1,|{\bf k}_{1T}|^2,\mu^2)
\Phi_\gamma(x_2,|{\bf k}_{2T}|^2,\mu^2) \frac{\overline{|{\cal
A}(R + R \to {\cal H})|^2}}{(2 x_1 x_2 y_1 y_2 S)^2},
\end{eqnarray}
where $x_1 = \xi_1/y_1$, $x_2 = \xi_2/y_2$, and ${\bf k}_{2T} = {\bf
p}_{T} - {\bf k}_{1T}$.
In the case of double-resolved photoproduction via the $2\to2$ subprocess,
we have
\begin{eqnarray}
&&\frac{d\sigma^{\mathrm{KT}}(e^+ + e^- \to e^+ + e^-+ {\cal H} +
X)}{d|{\bf p}_T|^2 d y}=\frac{1}{2 (2 \pi)^3}\int{d y_1} \int{d
y_2} \int{d|{\bf k}_{1T}|^2}\int{d \varphi_1}
\nonumber\\
&&\times
\int{dx_2} \int{d|{\bf k}_{2T}|^2} \int{d\varphi_2}\,
f_{\gamma/e}(y_1)\Phi_\gamma(x_1,|{\bf k}_{1T}|^2,\mu^2)
f_{\gamma/e}(y_2)\Phi_\gamma(x_2,|{\bf k}_{2T}|^2,\mu^2)
\nonumber\\
&&\times
\frac{y_2}{x_2 y_2 - \xi_2}\,
\frac{\overline{|{\cal A}(R + R \to {\cal H} + g)|^2}}{(2 x_1 x_2
y_1 y_2 S)^2},
\end{eqnarray}
where
\begin{equation}
x_1=\frac{1}{y_1 (x_2 y_2 - \xi_2) S}\left[({\bf k}_{1T} + {\bf
k}_{2T} - {\bf p}_{T})^2 - M^2 - |{\bf p}_{T}|^2 + x_2 y_2 \xi_1
S\right].
\end{equation}

\begin{figure}[ht]
\begin{center}
\includegraphics[width=1.0\textwidth, clip=]{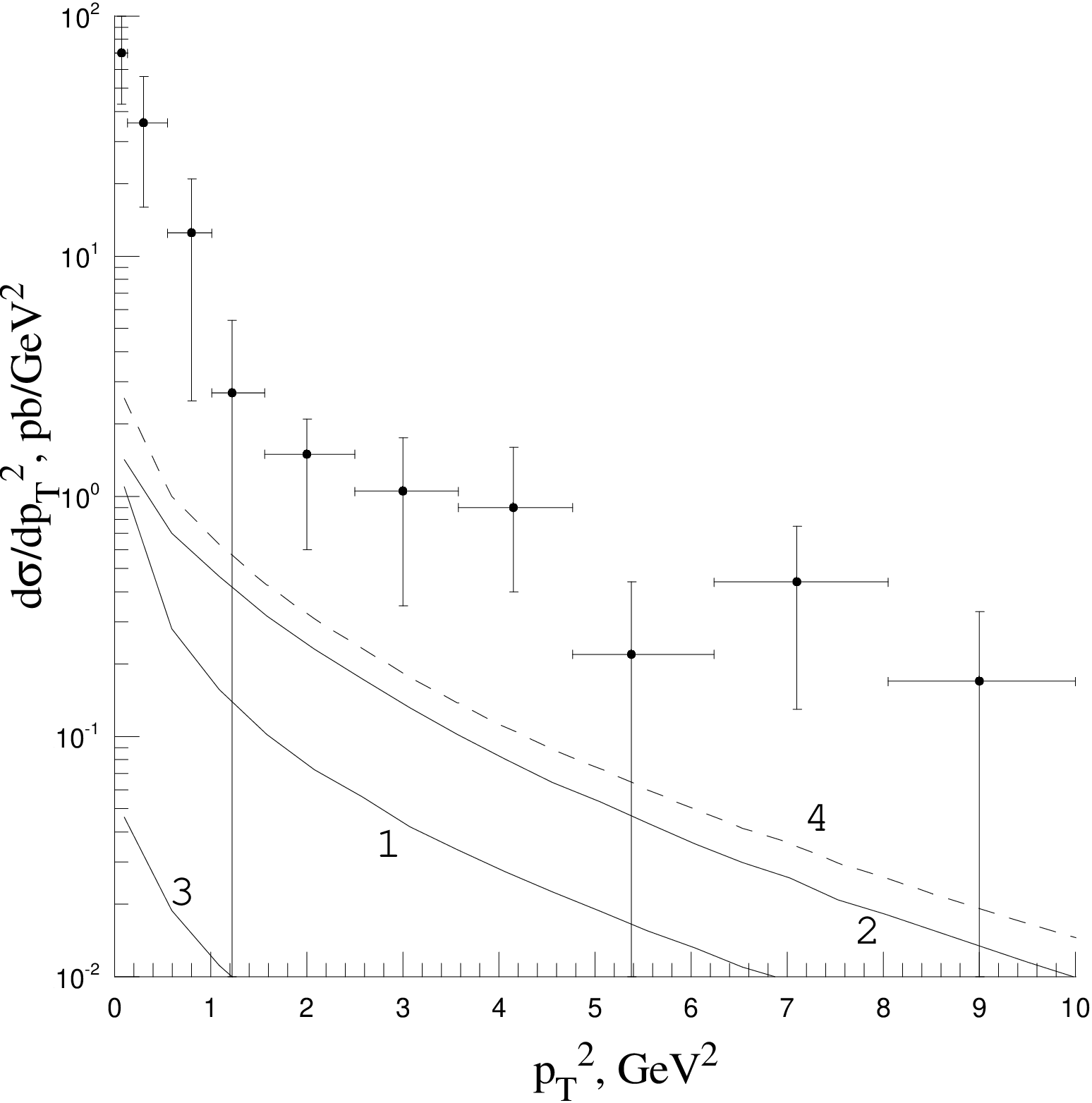}
\end{center}
\caption{\label{fig:LEP}Contributions to the $p_T^2$ distribution of prompt
$J/\psi$ photoproduction in $e^+e^-$ annihilation with $\sqrt{S}=197$~GeV,
$Q^2<9.93$~GeV$^2$, $W<35$~GeV, and $|y|<2$ from the partonic subprocesses
(1) $R + \gamma\rightarrow {\cal H}[{^1S}_0^{(8)},{^3P}_J^{(8)}]$,
(2) $R + \gamma\rightarrow {\cal H}[^{3}S_1^{(1)}]+g$,
(3) $\gamma + \gamma\rightarrow {\cal H}[^{3}S_1^{(8)}]+g$,
$R + R \to {\cal H}[{^3S}_1^{(8)}, {^1S}_0^{(8)}, {^3P}_{J}^{(8)}]$, and
$R + R \to {\cal H}[{^3S}_1^{(1)}] + g$, and
(4) their sum
compared with DELPHI data from LEP2 \cite{LEPJpsi}.
The theoretical results are obtained with the JB \cite{JB}
unintegrated gluon distribution function.}
\end{figure}

In Fig.~\ref{fig:LEP}, we confront the $p_T^2$ distribution of
$e^+ + e^- \to e^+ + e^- + J/\psi + X$, where $X$ is devoid of prompt photons,
measured by DELPHI \cite{LEPJpsi} with our full theoretical prediction (line
No.~4), which is broken down into the single-resolved color-octet contribution
(line No.~1), the single-resolved color-singlet contribution (line No.~2), and
the direct plus double-resolved contributions (line No.~3).
We observe that the single-resolved contribution makes up the bulk of the
cross section, while the direct and double-resolved contributions are greatly
suppressed, and that, within the single-resolved contribution, the
color-singlet channel is dominant.
The experimental data overshoot the theoretical prediction by a moderate
factor of 2--3.
For the case of $\gamma\gamma$ collisions, we conclude that the color-singlet
processes are dominant in the $k_T$-factorization approach, a situation
familiar from photo- and electroproduction in $ep$ collisions considered in
Sec.~\ref{sec:six}.
The situation is quite different for the collinear parton model, where
color-octet processes dominate \cite{KniehlLEP}.

Recently, in Ref.~\cite{ZotovLEP}, it was attempted to interpret the DELPHI
data in the $k_T$-factorization approach invoking only the CSM and neglecting
the cascade decays of the $\psi^\prime$ and $\chi_{cJ}$ mesons.
Curve No.~2 in Fig.~\ref{fig:LEP} approximately agrees with the corresponding
predictions in Ref.~\cite{ZotovLEP} for $m_c=1.55$~GeV.
In Ref.~\cite{ZotovLEP}, a significantly lower value of $m_c$ is employed to
reach agreement with the DELPHI data.

\section{\label{sec:eight}Conclusion}

Working at LO in the $k_T$-factorization approach to NRQCD, we analytically
evaluated the squared amplitudes of prompt charmonium production by reggeized
gluons in $RR$, $R\gamma$, and $Re$ collisions.
We extracted the relevant color-octet NMEs,
$\langle {\cal O}^{\cal H}[^3S_1^{(8)}]\rangle$,
$\langle {\cal O}^{\cal H}[^1S_0^{(8)}]\rangle$, and
$\langle {\cal O}^{\cal H}[^3P_0^{(8)}]\rangle$ for
${\cal H}=J/\psi$, $\psi^\prime$, and $\chi_{cJ}$ through fits to $p_T$
distributions measured by the CDF Collaboration in $p\bar p$ collisions at the
Tevatron with $\sqrt{S}=1.8$~TeV \cite{CDFI} and 1.96~TeV \cite{CDFII} using
three different versions of unintegrated gluon distribution function, namely
JB \cite{JB}, JS \cite{JS}, and KMR \cite{KMR}.
Appealing to the assumed NRQCD factorization, we used the NMEs thus obtained
to predict various cross section distributions of prompt $J/\psi$
photoproduction and electroproduction in $ep$ collisions and photoproduction
in $e^+e^-$ collisions and compared them with ZEUS \cite{epZEUS} and H1
\cite{epH1} data from HERA and DELPHI \cite{LEPJpsi} data from LEP2,
respectively.
In the case of photoproduction, we included both the direct and resolved
contributions.
As for the unintegrated parton distribution functions of the proton and the
resolved photon, we assumed the gluon content to be dominant.

Our fits to the Tevatron data turned out to be satisfactory, except for the
one to the $\chi_{cJ}$ sample based on the JB gluon density in the proton,
where the fit result significantly exceeded the measured cross section in the
small-$p_T$ region.
We found agreement with the HERA and LEP2 data within a factor of 2, which is
the typical size of the theoretical uncertainty due to the lack of knowledge
of the precise value of the $c$-quark mass and the NLO corrections.
Specifically, we found that direct and resolved photoproduction in $ep$
collisions under HERA kinematic conditions dominantly proceed through
color-singlet processes, namely $R(p)+\gamma \to {\cal H}[{^3}S_1^{(1)}] + g$
and $R(p) + R(\gamma) \to {\cal H}[{^3}S_1^{(1)}] + g$, respectively.
Similarly, photoproduction in $e^+e^-$ collisions under LEP2 kinematic
conditions is mainly mediated via the color-singlet subprocess
$R(\gamma) + \gamma \to {\cal H}[{^3}S_1^{(1)}] + g$, but the color-octet
subprocess $R(\gamma)+\gamma \to {\cal H}[{^1}S_0^{(8)}]$ also contributes
appreciably.

LO predictions in both the collinear parton model and the $k_T$-factorization
framework suffer from sizeable theoretical uncertainties, which are largely
due to unphysical-scale dependences.
Substantial improvement can only be achieved by performing full NLO analyses.
While the stage for the NLO NRQCD treatment of $2\to2$ processes has been set
in the collinear parton model \cite{kkms}, conceptual issues still remain to
be clarified in the $k_T$-factorization approach.
Since, at NLO, incoming partons can gain a finite $k_T$ kick through the
perturbative emission of partons, one expects that essential features produced
by the $k_T$-factorization approach at LO will thus automatically show up at
NLO in the collinear parton model.

\begin{acknowledgments}

V.A.S. and D.V.V. thank the 2nd Institute for Theoretical Physics at the
University of Hamburg for the hospitality extended to them during visits when
this research was carried out.
The work of D.V.V. was supported in part by a Mikhail Lomonosov grant, jointly
funded by DAAD and the Russian Ministry of Education, by the International
Center of Fundamental Physics in Moscow, and by the Dynastiya Foundation.
This work was supported in part by BMBF Grant No.\ 05~HT4GUA/4.

\end{acknowledgments}

\newpage

\begin{table}[hp]
\begin{center}
\caption{\label{tab:NME}NMEs for $J/\psi$, $\psi^\prime$, and $\chi_{cJ}$
mesons from fits in the collinear parton model (PM) \cite{BKLee} using the
MRST98LO parton distribution functions of the proton \cite{MRST} and in the
$k_T$-factorization approach using the JB \cite{JB}, JS \cite{JS}, and KMR
\cite{KMR} unintegrated gluon distribution functions.
The CDF prompt data from run~I \cite{CDFI} and run~II \cite{CDFII} have been
excluded from our fit based on the JB gluon density.}
\begin{ruledtabular}
\begin{tabular}{ccccc}
NME & PM \cite{BKLee} & Fit JB & Fit JS & Fit KMR \\
\hline $\langle {\cal O}^{J/\psi}[^3S_1^{(1)}]\rangle/$GeV$^3$ &
1.3 & 1.3 & 1.3 & 1.3 \\
$\langle {\cal O}^{J/\psi}[^3S_1^{(8)}]\rangle/$GeV$^3$ &
$4.4\times10^{-3}$ & $1.5\times10^{-3}$ & $6.1\times10^{-3}$ &
$2.7\times10^{-3}$ \\
$\langle{\cal O}^{J/\psi}[^1S_0^{(8)}]\rangle/$GeV$^3$ & --- &
$6.6\times10^{-3}$ & $9.0\times10^{-3}$ &
$1.4\times10^{-2}$ \\
$\langle {\cal O}^{J/\psi}[^3P_0^{(8)}]\rangle/$GeV$^5$ &
--- & 0 & 0 & 0 \\
$M_{3.4}^{J/\psi}/$GeV$^3$ & $8.7\times10^{-2}$ &
$6.6\times10^{-3}$ & $9.0\times10^{-3}$ & $1.4\times10^{-2}$
\\\hline $\langle {\cal
O}^{\psi^\prime}[^3S_1^{(1)}]\rangle/$GeV$^3$ & $6.5\times10^{-1}$
& $6.5\times10^{-1}$ & $6.5\times10^{-1}$ &
$6.5\times10^{-1}$ \\
$\langle {\cal O}^{\psi^\prime}[^3S_1^{(8)}]\rangle/$GeV$^3$ &
$4.2\times10^{-3}$ & $3.0\times10^{-4}$ & $1.5\times10^{-3}$ &
$8.3\times10^{-4}$ \\
$\langle{\cal O}^{\psi^\prime}[^1S_0^{(8)}]\rangle/$GeV$^3$ &
--- & 0 & 0 & 0 \\
$\langle {\cal O}^{\psi^\prime}[^3P_0^{(8)}]\rangle/$GeV$^5$ & ---
& 0 & 0 & 0 \\$M_{3.5}^{\psi^\prime}/$GeV$^3$ &
$1.3\times10^{-2}$ & 0 & 0 & 0 \\
\hline $\langle {\cal O}^{\chi_{c0}}[^3P_0^{(1)}]\rangle/$GeV$^5$
& $8.9\times10^{-2}$ & $8.9\times10^{-2}$ & $8.9\times10^{-2}$ &
$8.9\times10^{-2}$ \\
$\langle {\cal O}^{\chi_{c0}}[^3S_1^{(8)}]\rangle/$GeV$^3$ &
$2.3\times10^{-3}$ & 0 & $2.2\times10^{-4}$ & $4.7\times10^{-5}$ \\
\hline
$\chi^2/\mathrm{d.o.f}$ & ---  & 2.2 & 4.1 & 3.0 \\
\end{tabular}
\end{ruledtabular}
\end{center}
\end{table}

\end{document}